\documentclass[12pt,preprint,numberedappendix]{aastex}

\usepackage{emulateapj5}
\usepackage{epsf}

\newcommand{\tx}[1]{\textrm{#1}}

\newcommand{\kms}{km~$\tx{s}^{-1}$}
\newcommand{\kmsmpc}{km\,$\tx{s}^{-1}$\,Mpc$^{-1}$}
\newcommand{\dvp}{$R^{1/4}\,$}

\newcommand{\mlu}{M$_\odot$/L$_{B,\odot}$}
\newcommand{\ml}{M$_*$/L$_B$}

\newcommand{\bo} {{\bf B}}
\newcommand{\lo} {{\bf L}}
\newcommand{\ro} {{\bf H}}
\newcommand{\sv} {{\bf \hat{s}}}
\newcommand{\dv} {{\bf \hat{d}}}
\newcommand{\nv} {{\bf \hat{n}}}

\newenvironment{inlinefigure}{
\def\@captype{figure}
\noindent\begin{minipage}{0.999\linewidth}\begin{center}}
{\end{center}\end{minipage}\smallskip}

\newenvironment{inlinetable}{
\def\@captype{table}
\noindent\begin{minipage}{0.999\linewidth}\begin{center}}
{\end{center}\end{minipage}\smallskip}

\slugcomment{ApJ, in press}

\shorttitle{High-z LSD}
\shortauthors{Treu \& Koopmans}

\begin{document}

\title{Massive Dark-matter halos and Evolution of Early--type Galaxies to
$z$$\approx$1}

\footnotetext[1]{Based on observations collected at W.~M. Keck
Observatory, which is operated jointly by the California Institute of
Technology and the University of California, and with the NASA/ESA
Hubble Space Telescope, obtained at STScI, which is operated by AURA,
under NASA contract NAS5-26555.}

\author{Tommaso Treu$\!$\altaffilmark{2,3,5} \& L\'eon V.E. Koopmans$\!$\altaffilmark{4,6,7}}
\altaffiltext{2}{Department of Physics and Astronomy,
University of California at Los Angeles, Los Angeles, CA 90095; ttreu@astro.ucla.edu}
\altaffiltext{3}{Hubble Fellow} 
\altaffiltext{4}{Space Telescope Science Institute, 3700 San Martin Dr, Baltimore, MD 21218 }
\altaffiltext{5}{California Institute of Technology, 
Astronomy, mailcode 105--24, Pasadena, CA 91125}
\altaffiltext{6}{California Institute of Technology, 
Theoretical Astrophysics, mailcode 130--33, Pasadena, CA 91125}
\altaffiltext{7}{Current Adress: Kapteyn Astronomical Institute, P.O.Box 800, 
9700 AV Groningen, The Netherlands}

\begin{abstract}
The combination of gravitational lensing and stellar dynamics breaks
the mass-anisotropy degeneracy and provides stringent constraints on
the distribution of luminous and dark matter in early-type (E/S0)
galaxies out to $z\approx1$. We present new observations and models of
three lens systems (CFRS03.1077, HST14176+5226, HST15433+5352) and the
combined results from the five field E/S0 lens galaxies at $z \approx
0.5{\rm -}1.0$ analyzed as part of the Lenses Structure \& Dynamics
(LSD) Survey. Our main results are: (i) Constant mass-to-light ratio
models are ruled out at $>99\%$ CL for all five E/S0 galaxies,
consistent with the presence of massive and extended dark-matter
halos. The range of projected dark-matter mass fractions inside the
Einstein radius is $f_{\rm DM}$=0.37--0.72, or 0.15--0.65 inside the
effective radius $R_{\rm e}$ for isotropic models. (ii) The average
effective power-law slope of the total (luminous+dark; $\rho_{\rm
tot}\propto r^{-\gamma'}$) mass distribution is
$\langle\gamma'\rangle$=1.75$\pm$0.10~(1.57$\pm$0.16) for
Osipkov-Merritt models with anisotropy radius $r_i=\infty$($R_e$) with
an rms scatter of 0.2 (0.35), i.e.\ marginally flatter than isothermal
($\gamma'=2$). The ratio between the observed central stellar velocity
dispersion and that from the best-fit singular isothermal ellipsoid
(SIE) lens model is $\langle f_{\rm
SIE}\rangle=\langle\sigma/\sigma_{\rm SIE}\rangle=0.87\pm0.04$ with
0.08 rms, consistent with flatter-than-isothermal density
profiles. Considering that $\gamma'>2$ and $f_{\rm SIE}>1$ have been
reported for other systems (i.e.\ B1608+656 and PG1115+080), we
conclude that there is a significant intrinsic scatter in the slope of
the mass-density profile of lens galaxies (rms $\sim$15\%), similar to
what is found for local E/S0 galaxies. Hence, the isothermal
approximation is not sufficiently accurate for applications that
depend critically on the slope of the mass-density profile, such as
the measurement of the Hubble Constant from time-delays. (iii) The
average inner power-law slope $\gamma$ of the dark-matter halo is
constrained to be $\langle \gamma \rangle =1.3^{+0.2}_{-0.4}$ (68\%
C.L.), if the stellar velocity ellipsoid is isotropic ($r_i=\infty$)
or an upper limit of $\gamma<0.6$, if the galaxies are radially
anisotropic ($r_i=R_{\rm e}$). The observed range of slopes of the
inner dark-matter distribution is consistent with the results from
numerical simulations only for an isotropic velocity ellipsoid and if
baryonic collapse and star-formation do not steepen dark-matter
density profiles. (iv) The average stellar mass-to-light ratio evolves
as $d \log( M_*/L_{\rm B})/d z$=$-0.72\pm0.10$, obtained via a
Fundamental Plane analysis. An independent analysis based on lensing
and dynamics gives an average $\langle d \log( M_*/L_{\rm B})/d
z\rangle $=$-0.75\pm0.17$. Both values indicate that the mass-to-light
ratio evolution for our sample of field E/S0 galaxies is slighly
faster than those in clusters, consistent with the hypothesis that
field E/S0 galaxies experience secondary bursts ($\sim$10\% in mass)
of star formation at $z<1$.  These findings are consistent with pure
luminosity evolution of E/S0 galaxies in the past 8 Gyrs, and would be
hard to reconcile with scenarios involving significant structural and
dynamical evolution.

\end{abstract}

\section{Introduction}

A central assumption of the current standard $\Lambda$CDM cosmological
model is that galaxies form and evolve inside dark-matter halos (White
\& Rees 1978; Blumenthal et al.\ 1984; Davis et al.\ 1985). Dark
matter halos are ubiquitous and possibly universal in their density
profile (Navarro, Frenk \& White 1996,1997, hereafter NFW; Moore et
al.\ 1998) and they dominate the dynamics of large scale structures.
In spite of decades of searches and technological advances our
empirical knowledge of dark halos remains very sparse.

In the local Universe, dark-matter halos have been convincingly
detected -- predominantly through dynamical tracers -- in spiral
galaxies (e.g. Rubin, Thonnard, \& Ford 1978,1980; Faber \& Gallagher
1979; van Albada \& Sancisi 1986; Salucci \& Burkert 2000; Jimenez,
Verde \& Oh 2003), dwarf and low surface brightness galaxies (de Blok
\& McGaugh 1997; van den Bosch et al.\ 2000; Swaters, Madore \&
Trewella 2000), clusters of galaxies (Zwicky 1937; Kneib et al. 1993;
Lombardi et al.\ 2000; Sand et al. 2002, 2004; Kneib et al.\ 2003),
and -- at least in some cases -- in early-type type galaxies
(e.g. Fabbiano 1989; Mould et al. 1990; Matsushita et al. 1998;
Loewenstein \& White 1999; Saglia, Bertin \& Stiavelli 1992; Bertin et
al. 1994; Arnaboldi et al. 1996; Franx et al. 1994; Carollo et al.\
1995; Rix et al.\ 1997; Gerhard et al.\ 2001; Borriello, Salucci \&
Danese 2003; Seljak 2002). However, observational evidence regarding
dark-matter halos of early-type galaxies (E/S0s) is limited. In a
number of cases, constant mass-to-light models, with a mass-to-light
ratio consistent with those of a normal stellar population, appear
sufficient to explain the information available from mass tracers, and
there is no need to invoke the existence of dark-matter halos (e.g.\
Bertin et al.\ 1994; Romanowsky et al.\ 2003).

If detection of dark-matter is hard and often ambiguous, decomposing
the mass distribution into a luminous (mostly stellar) and dark-matter
component, to measure their relative contribution and spatially
resolved properties, has been possible in only very few cases with
varying results (see references above). The main hurdles to overcome
are the paucity of dynamical tracers at large radii (such as HI gas in
spirals) and the degeneracy between kinematic properties of dynamical
tracers (e.g.\ anisotropy for stellar dynamics) and the mass
distribution. For simplicity, we will refer to the latter problem as
the mass-anisotropy degeneracy.

The distant Universe ($z>0.1$) is an almost completely uncharted
territory. Gravitational lensing has provided evidence for a mass
distribution that is more extended than the stellar component, either
by the analysis of individual systems or by considering statistical
ensembles (e.g.\ Kochanek 1995, Rusin \& Ma 2001, Ma 2003, Rusin,
Kochanek \& Keeton 2003; Cohn et al. 2001, Mu\~{n}oz, Kochanek \&
Keeton 2001; Rusin et al.\ 2002; Winn, Rusin \& Kochanek 2003;
Wucknitz, Biggs \& Browne 2004). Unfortunately, for most lenses it has
proven very difficult to reliably separate the luminous from the
dark-matter component using lensing alone, and determine a precise
dark-matter density profile and mass fraction.

In spite of the difficulties, a detailed exploration of high-$z$
galaxies would come with a great reward, offering the opportunity to
map directly the evolution of dark and stellar mass over cosmic
time. By mapping the time (i.e.\ redshift) evolution of the relative
distributions of luminous and dark matter in early-type galaxies -- as
well as the evolution of stellar mass-to-light ratio and the slope of
dark-matter halos -- we can address directly the following
questions. How and when is mass assembled to form early-type galaxies?
What is their star formation history? Are dark-matter halos
characterized by cuspy mass profiles in the center as predicted by
numerical simulations? What is the role of star formation in shaping
the total mass distribution of early-type galaxies? Do isolated
early-type galaxies undergo internal structural/dynamical evolution?

\begin{figure*}[t]
\begin{center}
\leavevmode \hbox{ \epsfxsize=0.98\hsize \epsffile{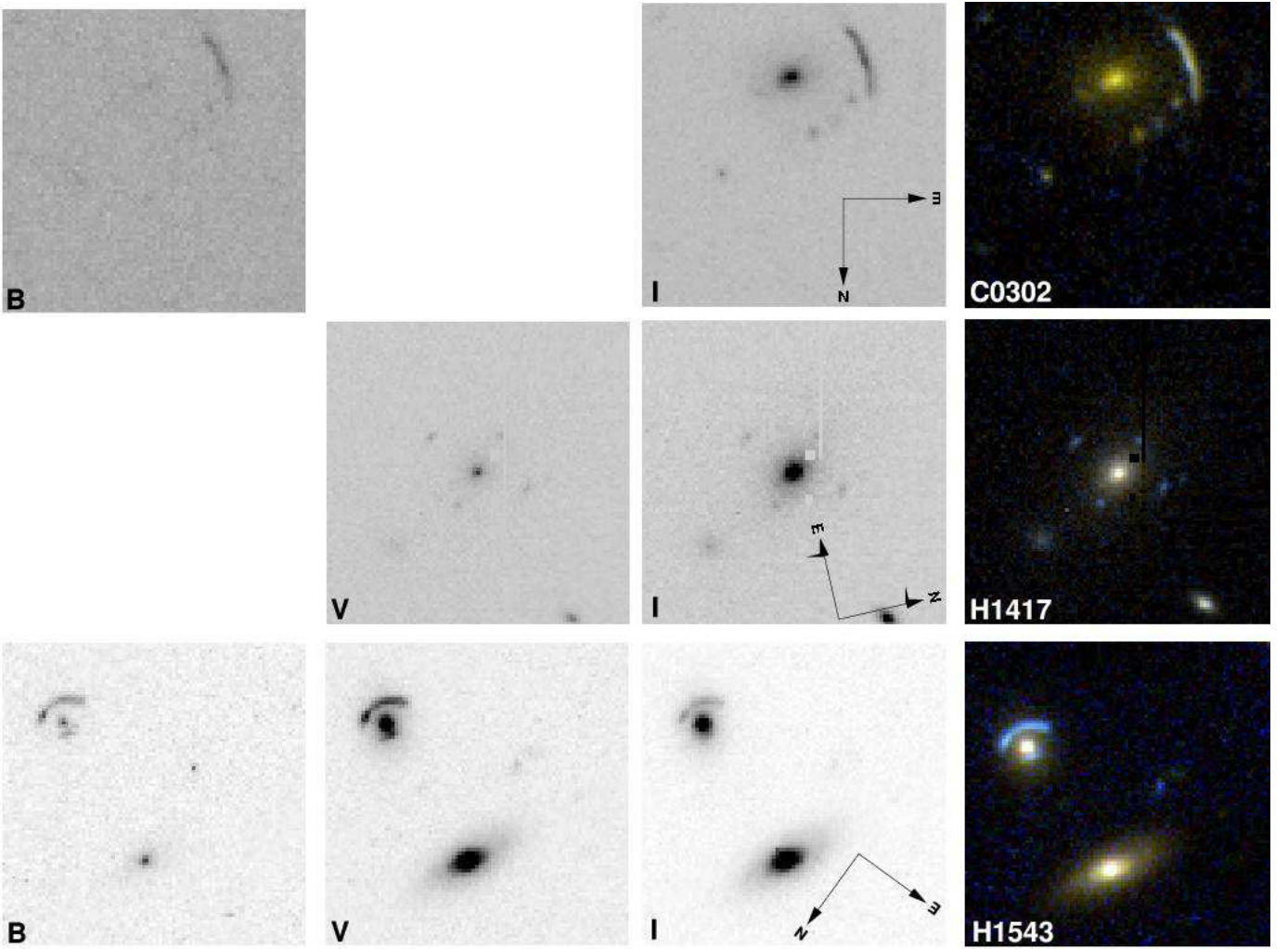} }
\figcaption{{\sl Hubble Space Telescope} (HST) WFPC2 images in B, V
and I band of the three gravitational-lens systems. The right panels
show a color-composite image.  The image sizes are 8$''$$\times$8$''$
for C0302 and 10$''$$\times$10$''$ for H1417 and H1543, and the
compass indicates the orientation of the images on the sky. Note the
bright galaxy G2 (z=0.506) next to the lens in H1543. C0302 is shown
off-center because it is close to the edge of the WFPC2 chip.
\label{fig:HSTimages}}
\end{center}
\end{figure*}

To answer these questions, we are undertaking the Lenses Structure and
Dynamics (LSD) Survey (Koopmans \& Treu 2002,2003; Treu \& Koopmans
2002a,2003; hereafter KT02, KT03, TK02a, TK03, or collectively KT).
The survey takes advantage of the fact that distant early-type
galaxies are efficient gravitational lenses. By focusing on lens
galaxies, we can use gravitational lensing analyses to provide a
precise and accurate mass measurement (typically 1--5 effective radii
R$_{\rm e}$), replacing very effectively the traditional dynamical
tracers at large radii (e.g.\ X-ray; planetary nebulae or globular
clusters kinematics) that cannot be used in the distant universe. The
lensing analysis is then combined with stellar kinematic measurements,
which provide constraints on the mass distribution at smaller radii
(typically $\la R_{\rm e}$). The combination of the two diagnostics
has proved to be very effective (KT) since they complement each other:
lensing provides a robust integrated mass measurement -- breaking the
mass-anisotropy degeneracy of the stellar dynamical analysis -- after
which stellar dynamics provides a handle on the mass density profile
of the lens.

The target lenses were selected from the sample of known galaxy-scale
systems (see e.g.\ the CASTLES web-page at URL
http://cfa-www.harvard.edu/castles/) for their morphology (E/S0),
brightness of the lens ($I\la 22$) and favorable contrast between the
lens and the source to allow for internal kinematic measurements, and
relative isolation (e.g.\ no rich clusters nearby) to simplify as much
as possible the lens model and reduce the related uncertainties.

Spectroscopic observations using the Keck Telescopes have now been
completed (a total of nine allocated nights between July 2001 and
December 2002), yielding exquisite internal kinematics for many
systems, including 9 early-type E/S0 lens galaxies in the range
$z\sim$\,0.1--1.0. In this paper, we present new data and models for
three lenses: CFRS.03.1077 (Crampton et al.\ 2002; Hammer et al. 1995;
Lilly et al.\ 1995; $z_l$=$0.938$, $z_s$=$2.941$ for the lens and
source respectively), HST1417+5226 (Ratnatunga et al.\ 1995; Crampton
et al. 1996; $z_l$=$0.810$, $z_s$=$3.399$), HST1543+5352 (Ratnatunga,
Griffiths \& Ostrander 1999a; the newly measured redshifts are
$z_l$=$0.497$, $z_s$=$2.092$). We will refer to these lenses as C0302,
H1417, H1543, respectively. Note that all three objects were
serendipitously discovered from HST images, i.e. the Groth Strip
Survey (Groth et al.\ 1994), the Medium Deep Survey (Ratnatunga,
Griffith \& Ostrander 1999b) and the HST Follow-up to the Canada
France redshift Survey (Brinchmann et al.\ 1998).  Together with the
analyses of the systems MG2016 and 0047 already presented by KT02,
TK02 and KT03, this completes the sample of the five high-redshift
($z\sim0.5-1.0$) pressure-supported systems targeted by the LSD Survey
so far. An analysis of the sample properties is presented here. The
data and analysis of the lower redshift systems and partly
rotationally supported systems in the current sample will be presented
in forthcoming papers.

The paper is organized as follows. In Section~\ref{sec:obs} we present
Hubble Space Telescope (HST) imaging and Keck spectroscopic
observations of the three lens systems. In Section~\ref{sec:FP} we use
the photometric and kinematic measurements to compare the sample of
E/S0 lens galaxies to the local Fundamental Plane (Djorgovski \& Davis
1987; Dressler et al. 1987). The offset from the local Fundamental
Plane is used to measure the redshift evolution of the stellar
mass-to-light ratio and thereby constrain their star formation
history. In Section~\ref{sec:lens} we present gravitational lens
models of the three lenses, using a modeling technique based on a
non-parametric source reconstruction (e.g.\ Wallington et al. 1996;
Warren \& Dye 2003) to fully take advantage of the extended nature of
the multiply imaged sources. In Section~\ref{sec:mass} we introduce
two-component (luminous plus dark-matter) mass models that will be
used in~Section~\ref{sec:analysis} to perform a joint lensing and
dynamics analysis and derive limits on the stellar mass-to-light
ratio, on the inner slope of the dark-matter halo and on the total
mass density profile of the sample of E/S0 galaxies.  We also consider
the complete sample of five high redshift lenses and discuss the
evolution of the stellar mass to light ratio in terms of stellar
population and structural/dynamical evolution, the dark-matter mass
fraction and the limits on the inner slope of the dark-matter halos
from a joint statistical analysis of the sample. In
Section~\ref{sec:homo} we discuss the homogeneity of the mass
distribution of the lens galaxy population, and its implication for
lens based studies such as the determination of the Hubble constant
from gravitational time delays, early-type galaxies evolution, and the
determination of cosmological parameters from lens statistics. A final
summary is given in~Section~\ref{sec:sum} and conclusions are drawn
in~Section~\ref{sec:conc}.

In the following, we assume that the Hubble constant, the matter
density, and the cosmological constant are
H$_0=65\,h_{65}$~km\,s$^{-1}$\,Mpc$^{-1}$ with $h_{65}=1$, $\Omega_{\rm
m}=0.3$, and $\Omega_{\Lambda}=0.7$, respectively. Throughout this
paper, $r$ is the radial coordinate in 3-D space, while $R$ is the
radial coordinate in 2-D projected space.

\begin{figure*}[t]
\begin{center}
\leavevmode \hbox{\epsfxsize=0.99\hsize \epsffile{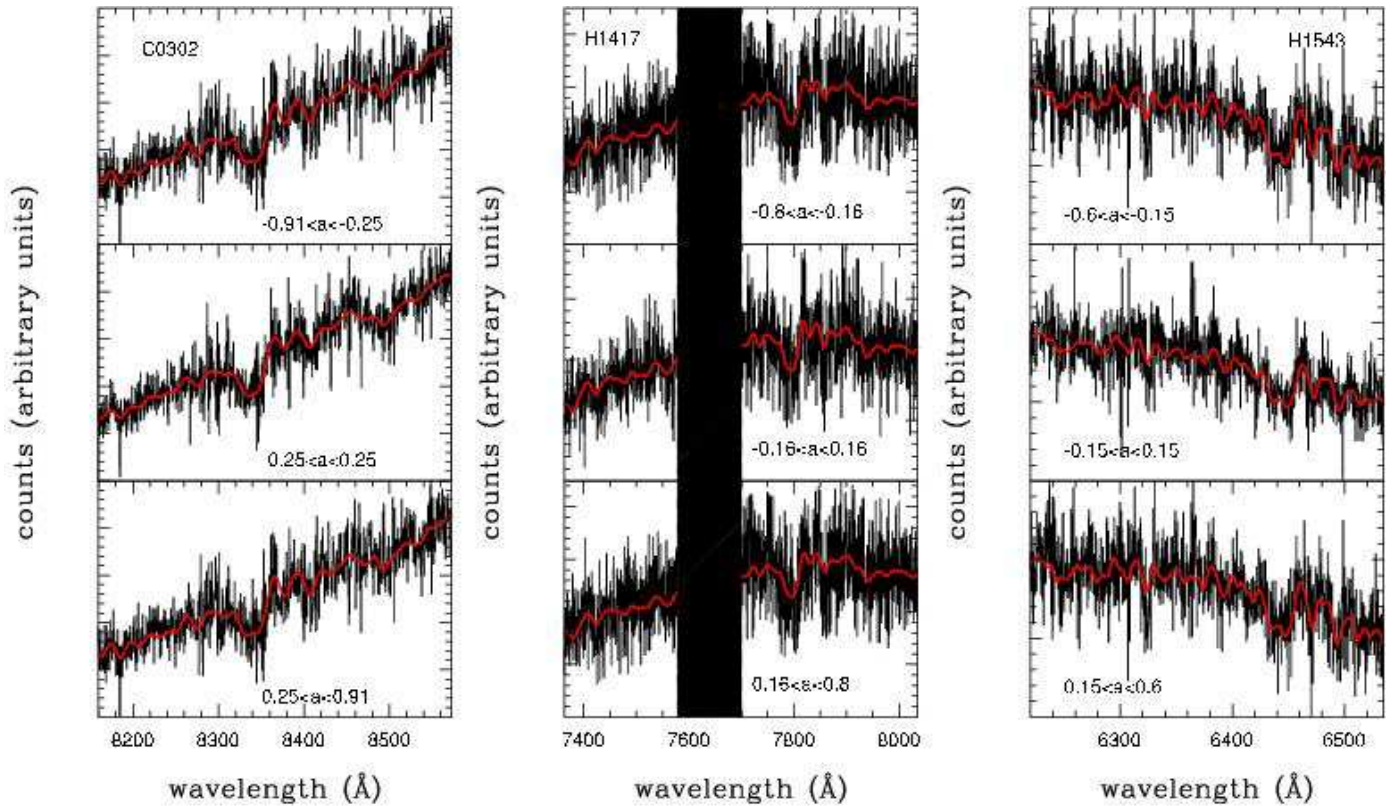}}
\figcaption{\label{fig:allspectra} Keck--ESI spectra of the E/S0 lens
galaxies in C0302, H1417 and H1543. A stellar template broadened to
the best fitting velocity dispersion is overplotted. The bin-sizes
along the major axes are indicated in arcseconds. The width of the
bins are $1\farcs25$.}
\end{center}
\end{figure*}

\section{Observations}

\label{sec:obs}

\subsection{HST imaging}

\label{ssec:photo}

Wide Field and Planetary Camera 2 (WFPC2) images of the systems
CFRS.03.1077, HST1417+5226 and HST1543+5352 are available from the HST
archive. Five exposures each through the F450W filters and the F814W
filter are available for C0302, with a total exposure time of 7000s and
6700s respectively. Four exposures (4400s) are available for H1417
through F814W. Sixteen exposures are available through filter F606W at
3 different Position Angles (PAs), with a total exposure time of
11200s. For HST1543, 3 exposures are available through filter F450W
(8800s), 3 through F606W (9000s) and 2 through F814W (6000s).
  
The images were reduced using a series of {\sc iraf} scripts based on
the {\sc iraf} package {\sc drizzle} (Fruchter \& Hook 2002), to align
the different pointings and perform cosmic ray rejection. The images
were combined on a $0\farcs1$ pixel scale. An exception was made for
the images of H1417 through F606W, which were combined in three groups
according to the HST PA, to avoid problems related to distortion
correction and complications of the azimuthal structure of the Point
Spread Function (PSF).  The reduced images of the galaxies are shown
in Figure~\ref{fig:HSTimages} together with color composite
images. Note that for H1417 only the F606W images at the same PA as
for the F814W ones are shown and that in these images a bad column of
WFPC2 runs very close to one of the multiple images, although this is
not the case for the other two sets of F606W images (not shown).

Surface photometry was performed on the F606W and F814W images as
described in Treu et al.\ (1999; hereafter T99) and Treu et al.\
(2001b; hereafter T01b). The F450W images were not used for surface
photometry given the low signal to noise of the (red) lens galaxies
and the large contamination from the (blue) multiple images. The
galaxy brightness profiles are well represented by an \dvp\ profile,
which we fit -- taking the HST point spread function (PSF) into
account -- to obtain the effective radius ($R_e$), the effective
surface brightness (SB$_e$), and the total magnitude.  The relevant
observational quantities of the lens galaxies and their errors are
listed in Table~\ref{tab:HST}. The errors on SB$_e$ and $R_e$ are
tightly correlated and that the uncertainty on the combination $\log
R_e - 0.32 SB_e$ that enters the Fundamental Plane (see
Section~\ref{sec:FP}) is very small ($\sim$0.015; see Kelson et
al. 2000; T01b; Bertin, Ciotti \& del Principe 2002).  The restframe
photometric quantities listed in Table~\ref{tab:HST} -- computed as
described in T01b -- are corrected for Galactic extinction using
E(B--V) from Schlegel, Finkbeiner \& Davis (1998).

Astrometry for the system H1417 -- the only one where the lensed
images can be approximated as point images -- was derived from the two
sets of 6 exposures through filter F606W that are not affected by a
bad column. The two astrometries agree within the uncertainties and
are averaged to determine the relative offsets between the lens galaxy
and the multiple images used to constrain the lens model in
Section~\ref{sec:lens}.

\medskip
\begin{table*}
\caption{Spectroscopic Observing log. \label{tab:log}}
\centering
\begin{tabular}{lccccc}
\hline
\hline
Galaxy & Instrument & Date & Seeing & Exptime & PA\\
\hline
C0302  & ESI        &  Dec 7,8 2002 &  $0\farcs8$      &  23400s     & 110\\
H1417  & ESI        &  Feb 7 2002   &  $0\farcs8$      &  6300s      & 37.2\\
H1417  & ESI        &  Jul 25,26 2001& $0\farcs7$      &  7200s      & 37.2\\
H1543  & ESI        &  Jun 6 2002   &  $0\farcs6$      &  14400s     & 68\\
H1543  & LRIS       &  Mar 5 2003   &  $1\farcs0$      &  5400s      & 142\\
\hline 
\end{tabular}
\end{table*}

\subsection{Keck Spectroscopy}

\label{ssec:spec}

The lens galaxies were observed using the Echelle Spectrograph and
Imager (ESI; Sheinis et al.\ 2002) at the Keck--II Telescope during
four runs on July 21--26 2001, February 7--8 2002, June 6 2002,
December 7--8 2002. Conditions were generally photometric with
episodes of thin cirrus. Between each exposure, we dithered along the
slit to allow for a better removal of sky residuals in the red end of
the spectrum.  The slit (20$''$ in length) was aligned with the major
axis of the lens galaxy (C0302, H1417) or slightly tilted as to
include the massive nearby companion (H1543;
Figure~\ref{fig:HSTimages}). ESI was used in high resolution mode with
a $1\farcs25$ wide slit, yielding a resolution of $\sim 30$ \kms,
adequate for measuring the stellar velocity dispersion and removing
narrow sky emission lines. The centering of the lens galaxies in the
slit was constantly monitored by means of the ESI viewing camera (all
galaxies were bright enough to be visible in a few seconds exposure)
and we estimate the centering perpendicular to the slit to be accurate
to $\la0\farcs1$.  Additional details of the observing runs are given
in Table~\ref{tab:log}.  The ESI data were reduced using the {\sc
iraf} package EASI2D\footnote{developed by D.~Sand and T.~Treu; Sand
et al. (2002,2004)} as described in KT.  The redshifts of the lenses are
given in Table~\ref{tab:HST}.

\begin{table*}
\caption{Observed spectro-photometric quantities \label{tab:HST}}
\begin{center}
\begin{tabular}{lccc}
\hline
\hline
Lens                             & C0302           & H1417           & H1543          \\   
\hline						                    
redshift (lens)                  & 0.938$\pm$0.001 & 0.810$\pm$0.001 & 0.497$\pm$0.001 \\
redshift (source)                & 2.941$\pm$0.001 & 3.399$\pm$0.001 & 2.092$\pm$0.001  \\
$F814W$ (mag)                    & 19.86$\pm$0.11  & 19.59$\pm$0.05  & 20.22$\pm$0.10   \\
$F606W$ (mag)                    & --               & 21.53$\pm$0.05  & 20.66$\pm$0.11   \\
SB$_{e,F814W}$ (mag/arcsec$^2$)  & 22.87$\pm$0.13  & 21.71$\pm$0.12  & 20.30$\pm$0.10   \\
SB$_{e,F606W}$ (mag/arcsec$^2$)  & --               & 24.06$\pm$0.14  & 21.80$\pm$0.11   \\
$R_{e,F814W}$ (arcsec)           & 1.60$\pm$0.15   & 1.06$\pm$0.08   & 0.41$\pm0.04$    \\ 
$R_{e,F606W}$ (arcsec)           & --               & 1.29$\pm$0.13   & 0.42$\pm0.04$    \\
$b/a$=$(1-e)$                    & 0.75$\pm$0.05   & 0.85$\pm$0.05   & 0.95$\pm$0.05    \\
Major axis P.A. ($^\circ$)       & $-$72$\pm5$     &   $34\pm 5$     &  56$\pm$5    \\
\hline						                     
$\sigma$  (km\,s$^{-1}$)         & 251$\pm$19      & 224$\pm$15      & 116$\pm$10       \\
$M_{V}-5\log h_{65}$ (mag)       & --               & --               & $-$21.40$\pm$0.10\\
$M_{B}-5\log h_{65}$ (mag)       & $-$23.30$\pm$0.1& -22.98$\pm$0.055& $-$20.63$\pm$0.11\\
$R_{e,V}$ ($h_{65}^{-1}$kpc)     & --               & --               & 2.7$\pm$0.3      \\
$R_{e,B}$ ($h_{65}^{-1}$kpc)     & 14.1$\pm$1.3    & 8.6$\pm$0.6     & 2.8$\pm$0.3      \\
SB$_{e,V}$  (mag/arcsec$^2$)     & --               &   --              & 19.34$\pm$0.10   \\
SB$_{e,B}$ (mag/arcsec$^2$)      & 20.92$\pm$0.14  & 20.25$\pm$0.12  & 20.16$\pm$0.11   \\
\hline
\hline
\end{tabular}
\end{center}
{\footnotesize Note: The second part of the table lists rest-frame
quantities, derived from the observed quantities as described in
Section~2. Note that $\sigma$ is the central velocity dispersion
corrected to a circular aperture of radius R$_e/8$.  All quantities in
this table assume H$_0=65$~km\,s$^{-1}$\,Mpc$^{-1}$, $\Omega_{\rm
m}=0.3$ and $\Omega_{\Lambda}=0.7$.}
\end{table*}

First, the full integrated spectra were used to derive the central
velocity dispersion of the lens galaxies with maximal accuracy. The
aperture velocity dispersions are measured by comparing broadened
stellar templates with the observed galaxy spectra in pixel space, as
described in Treu et al.\ (1999, 2001) and Koopmans \& Treu (2002,
2003). They are subsequently converted into a central velocity
dispersion (i.e.\ within a circular aperture of radius $R_e/8$),
applying an upward correction factor of 1.08$\pm$0.04 (Treu et al.\
2001). The central velocity dispersion of G2 -- the massive companion
to H1543 at z=0.506 (see Fig.~\ref{fig:HSTimages}) -- is found to be
$\sigma=263\pm11$ \kms, by applying the same procedure.

\begin{inlinefigure}
\begin{center}
\resizebox{\textwidth}{!}{\includegraphics{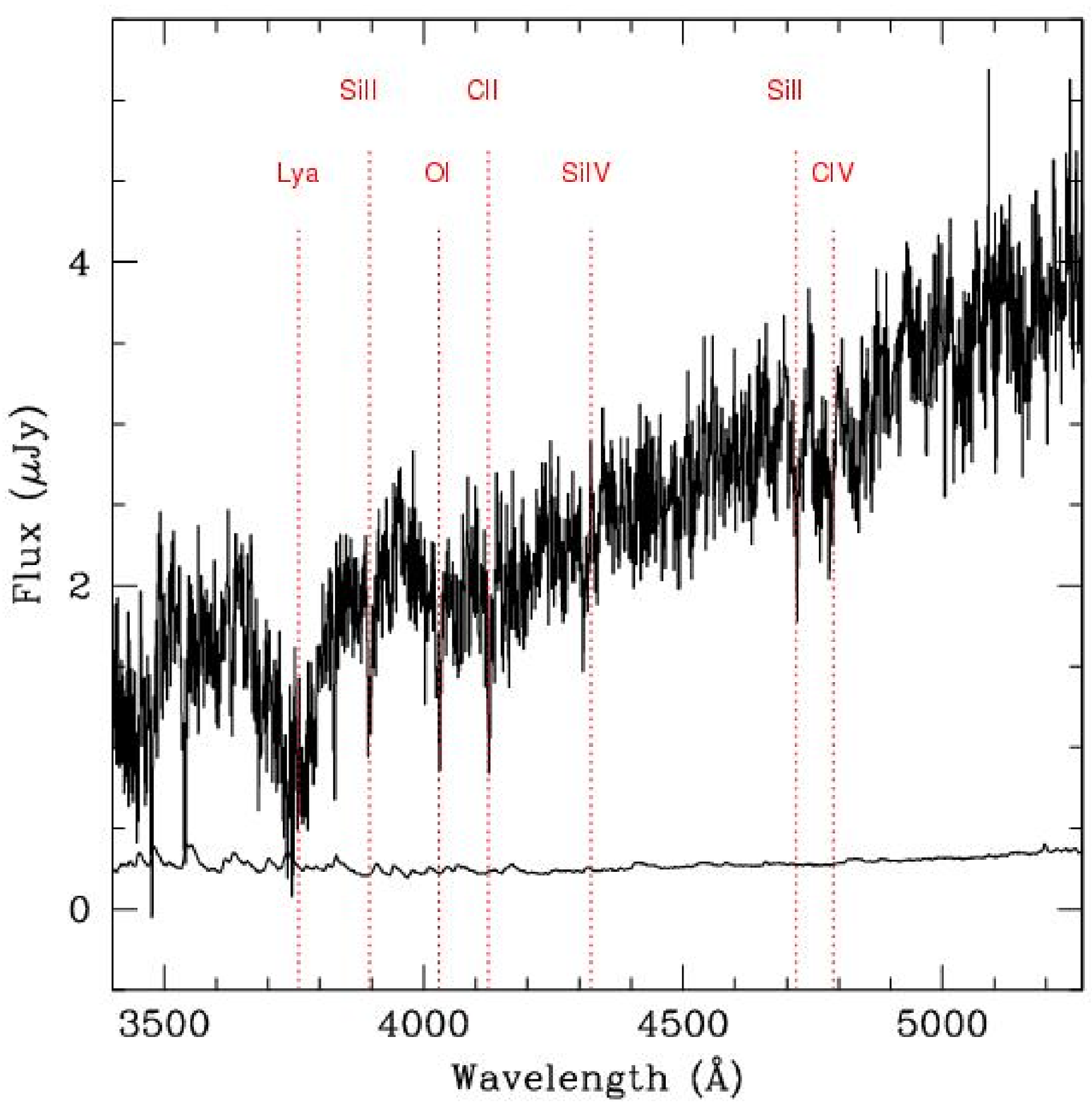}}
\end{center}
\figcaption{\label{fig:zH1543} A Keck--LRIS spectrum (3400--5300\,\AA)
of the lensed source in H1543. Multiple absorption features clearly
identify a redshift of $z_s=2.092$. The lower curve represents the
noise level.}
\end{inlinefigure}

Second, we derive spatially resolved kinematic profiles in the
following manner. For each galaxy, we define three symmetric
spectroscopic apertures centered on the lens galaxy, such that the
signal-to-noise ratio of the spectra integrated within each aperture
is sufficient to measure\footnote{Using the Gauss-Hermite Pixel
Fitting Software, van der Marel 1994, as described, e.g., in Treu et
al.\ 2001.} a stellar velocity dispersion (see the discussion in
Koopmans et al.\ 2003). Portions of the spectra of the lens galaxies
around the G-band (4304\AA) for each spectroscopic aperture are shown
in Figure~\ref{fig:allspectra}. The apertures are listed in
Table~\ref{tab:sigma} together with the measured stellar velocity
dispersions. Note that our measurement of the central velocity
dispersion of H1417, $\sigma=224\pm15$~\kms, is in excellent agreement
with the two measurements that have been published so far: Ohyama et
al.\ (2002) found $\sigma_{\rm ap}=230\pm14$~\kms\ within an aperture
equivalent to a circular aperture of radius $0\farcs4$; Gebhardt et
al.\ 2003 measure $\sigma_{\rm ap}=202\pm9$~\kms\ within their
spectroscopic aperture, which they correct to a central velocity
dispersion of $\sigma=222\pm10$~\kms.

Since the ESI spectra did not yield the redshift for the lensed arc in
H1543, additional spectra were obtained using LRIS (Oke et al.\ 1995)
on March 5 2003, to exploit the favorable contrast at the blue end of
the spectrum. The $1\farcs0$ slit was centered on the arc and aligned
with the parallactic angle (Table~\ref{tab:log}). The blue end of the
spectrum (Figure~\ref{fig:zH1543}) reveals a set of UV absorption
lines typical of Lyman-break galaxies (Steidel et al.\ 1996) that
unambiguously yields the redshift of the arc as $z_s{=}2.092\pm0.001$.

\section{The Fundamental Plane and the evolution of the stellar populations}

\label{sec:FP}

Early-type galaxies in the local Universe occupy approximately a plane
in the three-dimensional space defined by the parameters; effective
radius ($\log R_e$), effective surface brightness (SB$_e$) and central
velocity dispersion ($\log \sigma$),
\begin{equation}
\label{eq:FP} 
\log R_{\rm e} = \alpha_{\rm FP} \log~\sigma + \beta_{\rm FP}~{\rm SB}_{\rm
e} + \gamma_{\rm FP}
\end{equation} 
\noindent
known as the Fundamental Plane (hereafter FP; Dressler et al.\ 1987;
Djorgovski \& Davis 1987).

Under appropriate assumptions (e.g.\ Treu et al. 2001), the evolution
of the FP with redshift can be used to measure the star-formation
histories of early-type galaxies (e.g.\ Franx 1993). Specifically if
we can define an effective mass $M \propto \sigma^2 R_{\rm e}$, and if
there is no evolution of the slopes $\alpha_{\rm FP}$ and $\beta_{\rm
FP}$, the evolution of the intercept can be used to measure the
evolution of the average effective mass to light ratio $\Delta \log
M/L = - 0.4 \Delta \gamma_{\rm FP} / \beta_{\rm FP}$ (e.g.\ van Dokkum
\& Franx 1996; Kelson et al.\ 1997; Bender et al.\ 1998; Pahre 1998;
van Dokkum et al.\ 1998; J{\o}rgensen et al.\ 1999; Treu et al.\ 1999;
Ziegler et al.\ 2001; Treu et al.\ 2002; van Dokkum \& Stanford 2003;
van Dokkum \& Ellis 2003; Gebhardt et al.\ 2003; van der Wel et al.\
2004). In the following analysis we will adopt as a reference the
local FP from Bender et al.\ (1998), i.e. $\alpha=1.25$, $\beta=0.32$,
and $\gamma_{\rm FP}=-8.895-\log h_{50}$, noting that our results do
not depend critically on the adopted coefficients of the local FP
(e.g.  Treu et al.\ 1999, 2002).

\subsection{Stellar mass-to-light evolution from the FP}

\label{sec:mlevfp}

In Fig.\ref{fig:FP} we plot the evolution of the effective $M/L$ for
the five lens galaxies ($z\approx 0.5-1.0$) analyzed so far from the
LSD Survey (i.e.\ 0047$-$285, MG2016+112 and the three systems
discussed in Sect.~2), together with the published linear fits for the
FP $M/L$ evolution of cluster and field E/S0 galaxies.  The effective
$M/L$ evolution for the five lens galaxies is $d \log (M/L_{B})/dz
=-0.72\pm0.10$, i.e.\ E/S0 galaxies were on average brighter at $z=1$
by $1.82\pm0.26$ magnitudes in the restframe B band. 
This number
assumes that the intercept of the local FP of field E/S0 galaxies is
the same of that of Coma-Cluster galaxies, consistent with the very
small environmental dependence of the intercept in the local Universe
(less than 0.1 dex in $\log M/L_B$; e.g. van Dokkum et al. 2001;
Bernardi et al.\ 2003). For completeness, we mention that if the local
intercept is allowed to vary, the best fit values are $d \log
(M/L_{B})/dz =-0.42\pm0.19$ and
$\gamma_{FP,lenses}-\gamma_{FP,Coma}=-0.23\pm0.14$, where the error
bars are large because of the small number of points and the limited
redshift range covered by our sample. The full LSD sample -- including
the low redshift objects -- is needed to simultaneously fit for the
local intercept and its redshift evolution. In this paper we shall
therefore restrict our analysis to the fit of the redshift evolution.
We note that lens galaxies are not located in rich clusters and
therefore this sample of lens galaxies is more similar to the samples
of {\it field} E/S0s than those of cluster E/S0s.

In terms of the passive evolution of a single stellar population this
corresponds to a relatively recent epoch of formation $z_{\rm
ssp}\sim1.3$ (for a Bruzual \& Charlot GISSEL96 population synthesis
model with Salpeter IMF and solar metallicity; see Treu et al.\ 2001,
2002 for more details), i.e.\ somewhat younger stars than typically
observed for massive cluster E/S0 galaxies (for which $d \log
(M/L_{B}) / dz =-0.49\pm0.05$ corresponding to $z_{\rm ssp}\sim2$; van
Dokkum et al.\ 1998). However, this evolutionary rate is also
consistent with a scenario where most of the stars in field E/S0
galaxies are old and formed at $z>2$ (for example, MG2016 was found to
have a star-formation redshift of $\sim$2; KT02) , and the relatively
fast evolution is driven by secondary bursts of starformation
contributing $\sim10\%$ to the stellar mass between $z\sim1$ and
today. The latter scenario appears to be favored on the grounds of
three independent lines of evidence: (i) Early-type galaxies are
present in the field in significant numbers out to well beyond
$z\sim1$, inconsistent with a sudden creation at $z\sim1.3$ (e.g.
Treu \& Stiavelli 1999; Chen et al.\ 2002; Fukugita et al.\ 2004;
Glazebrook et al.\ 2004). (ii) Recent starformation in a fraction of
high-$z$ E/S0 galaxies is detected from alternative diagnostics, such
as colors (e.g.\ Menanteau et al.\ 2001; van de Ven et al.\ 2003) and
absorption and emission lines diagnostics (e.g. Treu et al.\ 2002;
Willis et al.\ 2002; van Dokkum \& Ellis 2003). (iii) A ``frosting''
of younger stars is found by comparing detailed stellar population
models with the spectra of local field E/S0s (e.g.\ Trager et al.\
2000).

The faster evolution of field versus cluster E/S0 galaxies is
qualitatively in agreement with the prediction of hierarchical models
(Kauffmann 1996; Diaferio et al. 2001; Benson et al. 2003), however
{\sl quantitatively} the difference is smaller than predicted (see
Treu 2004 for a recent review and discussion of this comparison from a
more general point of view.)

\begin{inlinefigure}
\begin{center}
\resizebox{\textwidth}{!}{\includegraphics{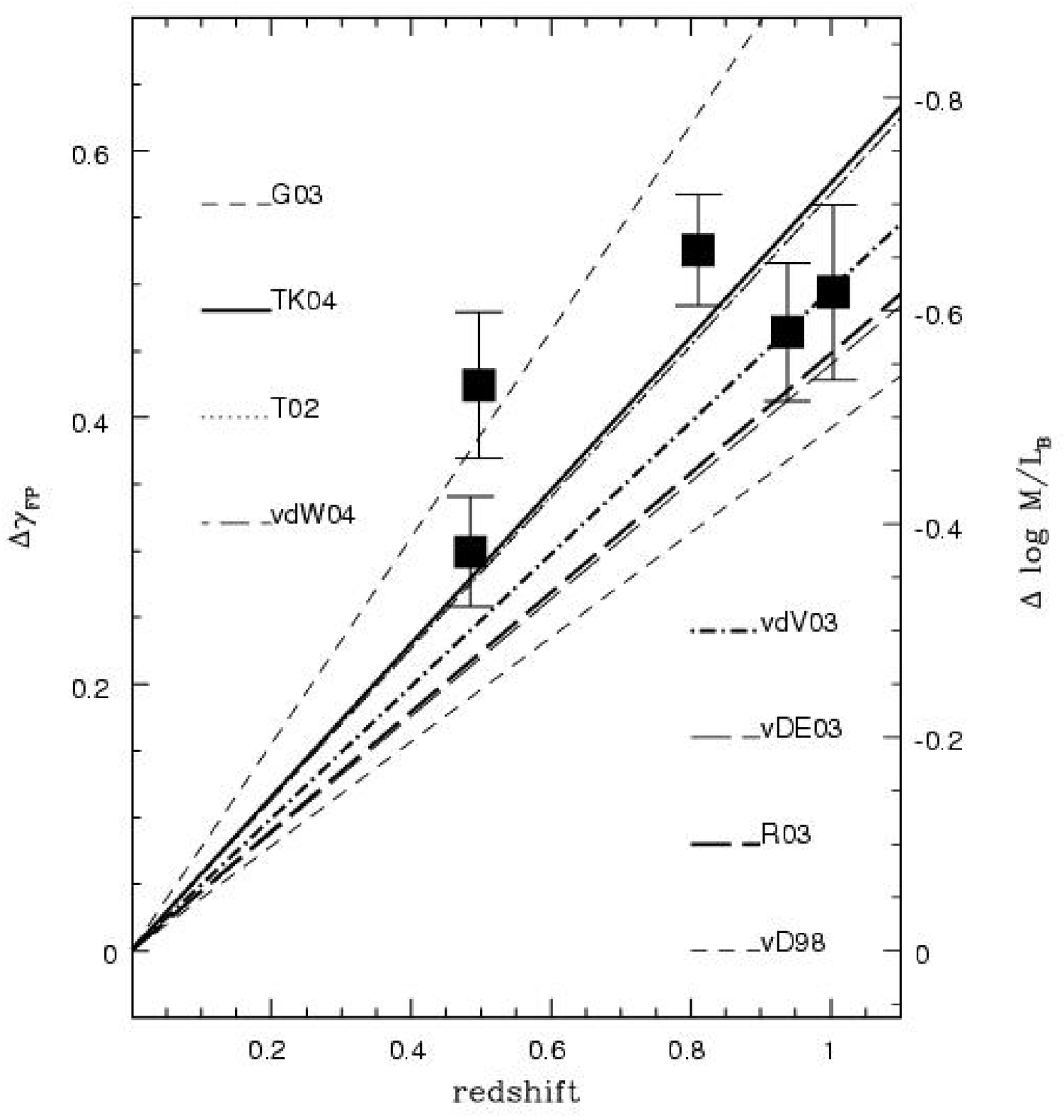}}
\end{center}
\figcaption{Evolution of the FP in rest frame B
band. Solid squares with error bars are the E/S0 lens galaxies
analysed in this paper. The heavy solid line (KT04) is a linear fit to
these points. The other lines represent linear fits to measurements
from the literature for the field FP (Treu et al.\ 2002=T02; Gebhardt
et al.\ 2003=G03; van Dokkum \& Ellis 2003=vDE03; van der Wel et al.\
2004=vdW04), the cluster FP (vD98=van Dokkum et al.\ 1998), and the FP
based on estimates of $\sigma_{\rm SIE}$ using the method of Kochanek
et al.\ 2000 (Rusin et al.\ 2003=R03; van de Ven et al.\
2003=vdV03). Note that the linear fits of T02, TK04 (this paper) and
vdW04 are almost identical. See text for details.
\label{fig:FP}}
\end{inlinefigure}

\subsection{Discussion and comparison with previous work}

Our measurement is in good agreement with the results -- using direct
measurements of the central stellar velocity dispersion, $\sigma$ --
published by Treu et al.\ (2002) who find $d \log (M/L_{B}) /
dz=-0.71^{+0.11}_{-0.16}$ and by van der Wel et al.\ (2004) who find
$-0.71\pm0.20$. In contrast, van Dokkum \& Ellis (2003) find a
marginally slower evolution $d \log (M/L_{B}) / dz= -0.55\pm0.05$,
while Gebhardt et al.\ (2003) measure a much faster evolution, i.e.\
2.4 magnitudes to $z=1$, corresponding to a value of $d \log
(M/L_{B})/dz \approx -0.96$.  At face value these results appear
inconsistent at the 1--2 $\sigma$ level.  Assuming that mutually
consistent measurement techniques have been adopted\footnote{As
suggested by the good agreement between the FP parameters measured by
different groups, see Section~\ref{ssec:spec} and Gebhardt et al.\
2003}, the differences could arise for a variety of reasons.

A first possible explanation is that the various samples cannot be
directly compared because of subtle differences between the
morphological classification schemes. Indeed van Dokkum \& Ellis
(2003) found that the 2 overluminous E/S0 galaxies in their sample of
9 showed asymmetric features in the Hubble Deep Field images. A larger
sample of objects with deep imaging is necessary to quantify this
effect.

Second, luminosity-selected samples favor overluminous objects and are
therefore biased toward faster evolution. The extent of the bias
depends on the intrinsic scatter of the FP and on the magnitude
limit. It can be corrected by taking into account the selection
procedure in the analysis (Treu et al.\ 2001, 2002). To further
complicate matters, the evolutionary rate could be a function of mass,
with more massive galaxies evolving slower than less massive ones
(e.g. van der Wel et al. 2004; note also that H1543, the lens galaxy
with the largest offset from the local FP is also the least massive
one), resulting in a change of the FP slopes with redshift. In this
case, the mean evolutionary rate would be a function of the mass range
of the sample (Treu et al. 2002). Unfortunately, luminosity selection
could also mimic a change in the slope, because less massive galaxies
would make the cut only if they are overluminous (Kelson et al.\
2000). 
We note here that a lens sample, although in principle
mass selected, could suffer from the same kind of bias because of the
effective magnitude limit imposed by spectroscopic follow-up. The
statistical analysis of larger samples of lens galaxies is essential
to determine simultaneously the evolution of the intercept, slopes and
scatter of the FP, correcting for selection effects.

\begin{table*}
\begin{center}
\caption{Kinematic data along the major axis of the lenses}
\label{tab:sigma}
\begin{tabular}{lcccc}
\hline
\hline
Galaxy     & Aperture              & $\sigma$& $\Delta\sigma$ & S/N\\
           &  ($\sq''$)               &(\kms)& (\kms)& (\AA$^{-1}$)\\
\hline
C0302      & ($-$0.91:$-$0.25)$\times$1.25 & 195  & 17 & 12 \\
           & ($-$0.25:$+$0.25)$\times$1.25 & 256  & 19 & 14 \\
           & ($+$0.25:$+$0.91)$\times$1.25 & 234  & 23 & 11 \\
\hline
H1417      & ($-$0.80:$-$0.16)$\times$1.25 & 223  & 20 & 13 \\
           & ($-$0.16:$+$0.16)$\times$1.25 & 212  & 18 & 15 \\
           & ($+$0.16:$+$0.80)$\times$1.25 & 199  & 22 & 12 \\
\hline
H1543      & ($-$0.60:$-$0.15)$\times$1.25 & 77   & 14 & 12 \\
           & ($-$0.15:$+$0.15)$\times$1.25 & 108  & 14 & 13 \\
           & ($+$0.15:$+$0.60)$\times$1.25 & 124  & 19 & 12 \\
\hline
H1543 (G2) & ($-$0.38:$+$0.38)$\times$1.25 & 253 & 10 & 25 \\
\hline
\end{tabular}
\end{center}
{\footnotesize Note: The adjacent rectangular apertures are indicated,
as well as the measured aperture velocity dispersions ($\sigma$),
their uncertainty ($\Delta \sigma$), and the average S/N per \AA\ in
the region used for the kinematic fit.}
\end{table*}

A third possible explanation is cosmic variance. If indeed the
secondary burst scenario is correct, at any given time the majority of
E/S0 galaxies would be observed to follow a quiescent passive
evolution path in the FP space, while a fraction of E/S0 galaxies
would be observed within 1--2 Gyrs after the secondary burst, while
overluminous. The fraction of overluminous E/S0 galaxies would depend
on the duty-cycle of secondary bursts. If, for example, $\sim$10\% of
the stellar mass is formed in each secondary burst, each galaxy undergoes
one secondary burst between $z=1$ and $z=0$, and the bursts are
detectable for 2 Gyrs, then a quarter of the E/S0 galaxies between
$0<z<1$ would be overluminous. Therefore, only a handful of
overluminous galaxies would be observed in the current samples of
galaxies and small number statistics could be dominating the
uncertainties.

Following Kochanek et al.\ (2000), both Rusin et al.\ (2003) and van
de Ven et al.\ (2003) recently used image separations (or Einstein
Radii) of arcsecond-scale strong lens systems to estimate the central
velocity dispersion of E/S0 lens galaxies and construct a FP. The key
assumption is that lens galaxies have isothermal mass profiles
(i.e. $\rho\propto r^{-2}$) and therefore the central velocity
dispersion can be obtained directly from the image separation with the
assumption that $\sigma\approx \sigma_{\rm SIE}$ (see Kochanek et al.\
2000). Under this assumption, Rusin et al.\ (2003) find $d \log
(M/L_{B}) / dz = -0.56\pm0.04$, and van de Ven et al.\ (2003) -- using
the photometry and image separations from Rusin et al.\ (2003) -- find
$d \log (M/L_{B}) / dz =-0.62\pm0.13$, using larger errors and a
different weighting scheme. The fact that these estimates are so close
to the direct measurements discussed above is indeed remarkable. Not
only do many of the above arguments, related to selection effects and
small sample statistics, also apply to lens galaxy samples, departures
from isothermal mass profiles, or effects such as a mass-sheet
degeneracy, introduce additional sources of uncertainty. In other
words, the relatively good agreement between the direct and indirect
methods tells us that the isothermal approximation is probably not
dramatically wrong.

In the next sections, via a joint lensing and dynamics analysis, we
will further examine the accuracy of this approximation. However,
before we proceed to the full analysis we can gain some insight by
looking at the three objects in our sample that are also in R03 and
vdV03: 0047, H1417, MG2016. The offset from the local FP are in
agreement within the errors for 0047 and MG2016 (see also R03), while
for H1417 we measure $\sim 0.1$ dex more evolution for $\log (M/L_{\rm
B})$. The difference is entirely due to the difference between our
{\sl direct} measurement of the stellar velocity dispersion
($\sigma=224\pm15$\kms) and the velocity dispersion inferred from the
image separation ($\sim$290\,\kms; vdV03, R03; next Section). Hence,
in at least one case, the image separation underestimates the stellar
velocity dispersion.  As we will show in Section 7, this is the
case in three out of five lens systems in our sample.

\section{Gravitational lens models}

\label{sec:lens}

As in previous work on LSD lens systems (KT), we model the three
systems (C0302, H1417 and H1543) with a Singular Isothermal Ellipsoid
(SIE) mass distribution:
\begin{equation}\label{eq:kappa}
  \kappa_{\rm SIE}(x,y)=\frac{b_l\,\sqrt{q}}{2 \sqrt{y'^2 + q^2 x'^2}},
\end{equation}
with $b_l=4\pi (\sigma_{\rm SIE}/c)^2 (D_d\,D_{ds}/D_s)$ and the major
axis aligned north-south (Kormann et al.\ 1994).  We define $(x',y')$
(in radians) in Eq.\ref{eq:kappa}\ as the frame centered on the lens
$(x_l,y_l)$ and aligned with the PA of the lens $\theta_l$.  Because
the mass enclosed by the elliptical critical curve is independent of
the flattening of the mass distribution $q=(b/a)$, we find that
$R_{\rm Einst}=D_d\,b_l$ corresponds to the equivalent Singular
Isothermal Sphere (SIS) Einstein radius. The equivalent SIS mass
($M_{\rm Einst}$) is that enclosed by the critical curve. Both $R_{\rm
Einst}$ and $M_{\rm Einst}$ are needed in the joint lensing and
dynamical analysis.  We refer to KT03 for additional discussion of the
models.  We note that the enclosed mass $M_{\rm Einst}$ is nearly
independent of the choice of mass model and depends predominantly on
the monopole of the lens potential (Kochanek 1991).  We also allow for
an external shear with strength $\gamma_{\rm ext}$ and position angle
$\theta_{\rm ext}$, corresponding to the potential $\psi_{\rm
ext}(x,y)=-R^2\,(\gamma_{\rm ext}/2)\,\cos2(\theta-\theta_{\rm ext})$,
where we define $R^2={(x-x_l)^2+(y-y_l)^2}$ to be the square of the
distance to the lens center (see Keeton et al. 2002). 
We emphasize that our lens model is fully two-dimensional and
therefore takes correctly into account deviations from circular
symmetry (e.g. Sand et al. 2003, Dalal \& Keeton 2003, Bartelmann \&
Meneghetti 2003).

Only H1417 has well-defined lensed images and fluxes
(Fig.~\ref{fig:HSTimages}). For this lens system, previously modeled
by Knudson, Ratnatunga \& Griffiths (2001), we use the `traditional'
modeling technique where the source is represented by a single point
in the source plane with a given flux, which is magnified and mapped
onto the lensed images in the image plane. We use the code described
in Koopmans et al. (2003). To model the extended images of C0302 and
H1543, we implemented a code (see Appendix) that incorporates
techniques described in Wallington et al. (1996) and Warren \& Dye
(2003), allowing the full use of all the lensing information contained
in the images.

\subsection{HST14176+5226}

We first model the lens system with a SIE mass distribution and
external shear only. The lens strength ($b_l$), centroid $(x_l,y_l)$,
ellipticity ($q$) and position angle ($\theta_l$) are free parameters,
as well as the external shear, the source position and its flux. The
lens-centroid and position angle are further constrained by the
observational priors given in Tables \ref{tab:HST} and
\ref{tab:astrom1417}. The best model has $\chi^2/{\rm NDF}=38$ for
NDF=4, using the image constraints in Table~\ref{tab:astrom1417}, and
is clearly not satisfactory\footnote{
The $\chi^2$ value is
dominated by the lens position ($\chi^2_{\rm l}$=84) and the lensed
image positions ($\chi^2_{\rm i}$=65). The flux-ratios contribute only
$\chi^2_{\rm r}$=2. Since the images are somewhat extended (see
Fig.~\ref{fig:HSTimages}) and have no apparent compact structure that
could be affected by stars (i.e.\ microlensing) or substructure, one
would not expect significant flux-ratio anomalies.}.

The next order of observable complexity that can be introduced is a
linear gradient in the surface density of the mass model and has a
potential of $\psi_{\rm g}(x,y)=(R^3/4)\,|\nabla \kappa_{\rm g}|
\,\cos(\theta - \theta_g)$ (see Keeton 2002). This gradient can be the
result of an external perturber (e.g.\ group or cluster), but also an
internal asymmetry of the lens (e.g.\ M=1 mode). We assume
$\kappa_{\rm g}=0$ for a line through the lens centroid,
i.e. $R^2={(x-x_l)^2+(y-y_l)^2}$, such that the convergence gradient
adds no mass inside a symmetric aperture on the lens centroid
$(x_l,y_l)$. This adds two more free parameters (i.e. $|\nabla
\kappa_{\rm g}|$ and $\theta_g$).  The best model in this case has
$\chi^2/{\rm NDF}=3.2$ for NDF=2, which is considerably better than a
model without a convergence gradient. We feel that adding more free
parameters is no longer justified, since both $\chi^2$ and NDF are
small.

\begin{figure*}
\begin{center}
\leavevmode \hbox{ \epsfxsize=0.97\hsize \epsffile{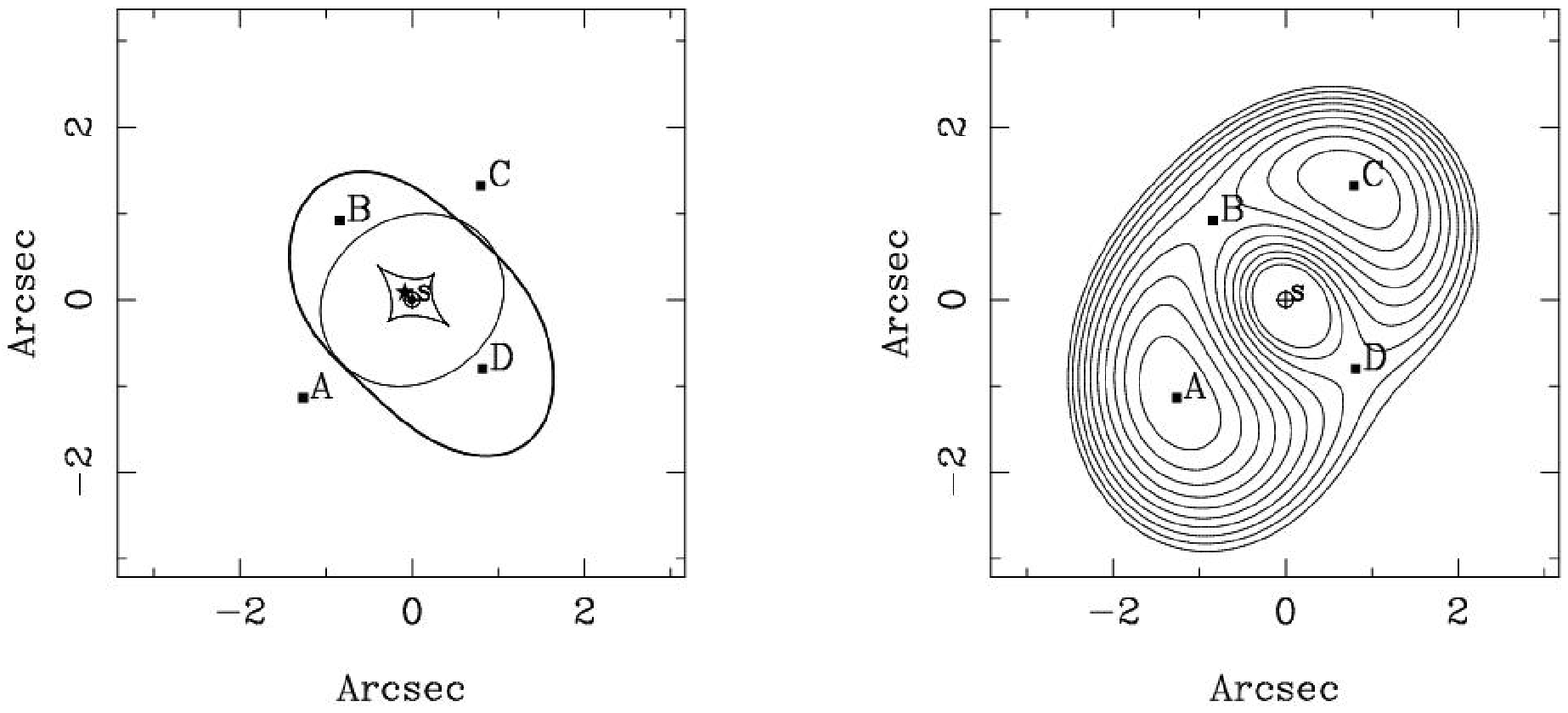} }
\figcaption{Best-fitting SIE lens model of H1417. The left panel shows
the position of the lens (solid circle) and the multiple images (solid
squares) on the image plane, together with the critical line (thick
solid line). The caustics (thin lines) and source position (solid
star) on the source plane are also shown. The right panel shows the
time delay surface with constant time-delay contours increasing from
0.0 (at A) to 83 days in steps of 8.3 days.\label{fig:lensmodel1417}}
\end{center}
\end{figure*}

\begin{inlinetable}
\tablecaption{Astrometry of the lens system H1417}
\begin{center}
\begin{tabular}{lccc}
  \hline
  Object & $\Delta$RA & $\Delta$ DEC & r \\
  \hline
  A    &   $-$1.266$\pm$0.006       &   $-$1.139$\pm$0.006   & 0.81$\pm$0.16\\ 
  B    &   $-$0.843$\pm$0.006	    &   $+$0.918$\pm$0.006   & 0.65$\pm$0.08\\
  C    &   $+$0.792$\pm$0.006	    &   $+$1.321$\pm$0.006   & 1.00$\pm$0.13\\
  D    &   $+$0.814$\pm$0.006	    &   $-$0.803$\pm$0.006   & 0.57$\pm$0.18\\ 
  G    &   $+$0.000$\pm$0.004	    &   $+$0.000$\pm$0.004   & -- \\
 \hline
\end{tabular}
\end{center}
\label{tab:astrom1417}
{\footnotesize Note: The column $r$ lists the F814W flux ratios of the
multiple images normalized to image C, taken from the CASTLeS 
web database.}
\end{inlinetable}

We notice that the gradient points in the direction of the major axis
of the SIE mass distribution and {\sl not} in the external shear
direction. The direction of the gradient and the major axis of the SIE
coincide within 11$^\circ$ and are consistent at the 2.5\,$\sigma$
level. We also notice that the agreement improves to within
$\sim$9$^\circ$ if the major axis is not constrained by a prior.  To
further examine this possible alignment, we also tested the SIE plus
shear and gradient model on PG1115+080 (a model similar to TK02b). We
find that for the best model, the position angle of the gradient and
external shear agree, both pointing to the confirmed external
perturber, which is a compact group $\sim$13$''$ from the lens (Kundi\v{c}
et al. 1997). In the case of H1417, however, there exists no obvious
group or cluster in the $\sim1'$ field around the lens
system\footnote{Although an overdensity of galaxies at $z\sim0.8$ in
the Groth Strip might be present; Koo et al.\ 1996; Im et al.\ 2002.}
and the gradient and shear position angle differ at the 12\,$\sigma$
level, nor are any of the galaxies around the lens massive enough to
account for the observed gradient of $|\nabla \kappa_{\rm
g}|=0.102\pm0.015$. A simple argument based on a SIS perturber shows
that one would expect its distance from H1417 to be related to the
Einstein radius of the perturber by $(R/R_{\rm Einst})^2\approx
1/(2|\nabla \kappa_{\rm g}|)$, or $R\approx 2.2 R_{\rm Einst}$. A
massive cluster with $R_{\rm Einst}=30''$ would still be detectable in
the field around the lens.  A single massive galaxy with $R_{\rm
Einst}=2''$ would be within $R\approx5''$ from the lens -- in the
direction of increasing $\kappa_{\rm g}$, i.e. $\theta_{\rm g}\approx
200^\circ$. Neither is obviously found. We therefore conclude that the
detected gradient is most likely associated with an asymmetry in the
lens mass distribution. A significant gradient is
also required for models with density slopes other than isothermal.

The best SIE model is shown in Fig.~\ref{fig:lensmodel1417} and all
model parameters are listed in Table~\ref{tab:astrom1417}. The
equivalent SIS velocity dispersion, mass and Einstein radius of the
lens are, $\sigma_{\rm SIE}=290\pm8$~\kms, $M_{\rm
Einst}=(70.8\pm7.6)\cdot 10^{10}$~M$_\odot$ and $R_{\rm
Einst}=11.4\pm0.6$~kpc (1\farcs41$\pm$0\farcs08), respectively (68\%
CL errors). Our derived Einstein radius (1\farcs41) agrees to within
$\le$0\farcs01 with that derived by Rusin et al. (2003). As noted the
gradient adds no mass. 

To further test the robustness of the enclosed mass measurement, we
examine three more sources of uncertainty. First, we find that the
mass enclosed by a circular aperture with 1\farcs41 radius, gives a
mass that is only 2.0\% lower than our best estimate of the mass
within the elliptical critical curve. Second, the best $\chi^2$ models
with density slopes that are 40\% steeper/shallower than isothermal
have enclosed masses different by only $-$2.0\%/+1.3\%. Third, if
$\kappa_{\rm g}$ is caused by an external perturber, a mass-sheet
($\kappa_{\rm sheet}$) must be associated with it, since truly
negative values of $\kappa_{\rm g}$ are not allowed. An estimate of
the external convergence can be obtained from the external shear,
assuming an isothermal mass distribution, i.e. $\kappa_{\rm
sheet}\sim\gamma_{\rm ext}$. If we set $\kappa_{\rm sheet}=\gamma_{\rm
ext}$, we find that the velocity dispersion, enclosed mass and
Einstein radius decrease by 2.5\%, 10\% and 5\%, respectively.
However, in the absence of any evidence for external perturbers that
could result in a non-local convergence, we conclude that the mass
measurement is robust, reliable and does not introduce a bias in the
final lensing plus dynamics analysis (Section~6). To account for most
of the uncertainties we assume a total 1--$\sigma$ errors of 10\% on
the enclosed mass.

\begin{figure*}
\begin{center}
\leavevmode \hbox{ \epsfxsize=0.95\hsize \epsffile{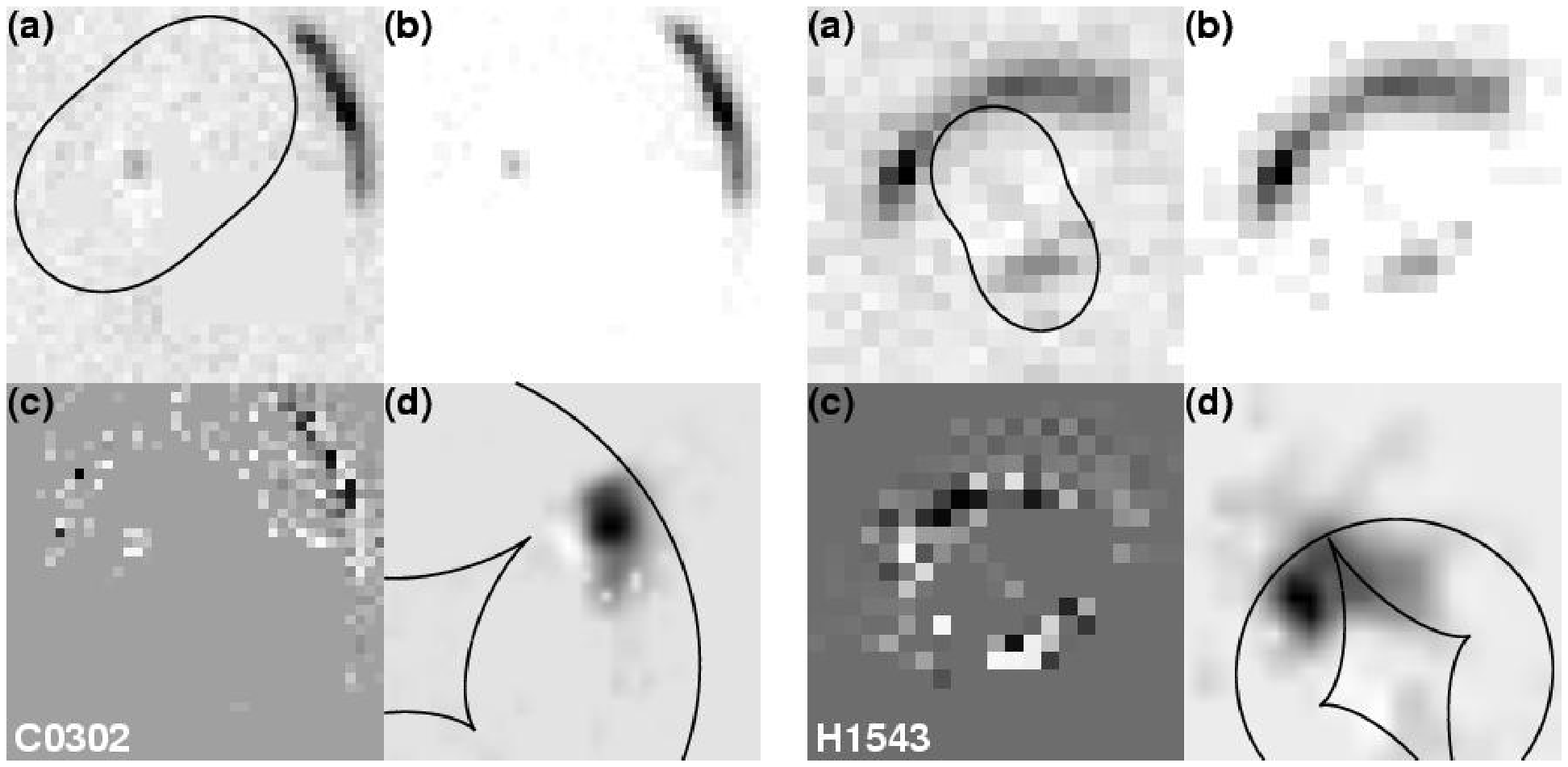}}
\figcaption{\label{fig:gridlens} The best {\tt LENSGRID}
(Section~\ref{sec:lens}) reconstructions of the extended arcs and
counter images in C0302 (left) and H1543 (right). Panels (a) show the
original image with lens-galaxy subtracted.  Panels (b) show the best
reconstruction. Panels (c) show the difference between observation and
model and panels (d) show the source in the source plane, regridded
and smoothed by a Gaussian with a FWHM=0\farcs08 to highlight its
structure. The curves are the critical (upper-left) and caustic
(lower-left) curves, respectively. }
\end{center}
\end{figure*}

\subsection{CFRS03.1077}

The lensed arc and counter image in the system C0302 (Crampton et al.\
2002; Hammer et al. 1995; Lilly et al.\ 1995) do not have enough
structure to allow a simple one-to-one mapping between them. To
determine the total enclosed mass, we implemented a lens code that
combines elements from Wallington et al. (1996; W96 from hereon) and
Warren \& Dye (2003; WD03) and can model extended lensed images on a
grid (i.e.\ CCD image). In the Appendix we outline the general
features of the code and where it differs from W96 and WD03.

In the discovery paper, Crampton et al. (2002) modeled the lens using
an NIE mass model with non-zero core radius. They find a good fit to
the arc and counter image, using an axisymmetric source model just
outside the cusp. A high velocity dispersion of 387$\pm$5~\kms\ is
inferred from their lens model, leading to the conclusion that the
lens galaxy is as faint as present-day ellipticals of similar central
velocity dispersion and thus shows no sign of passive evolution. Since
there are several dwarf-like galaxies near the main lens galaxy
(Fig.~1), possibly indicating the presence of a small group, the
apparent underluminous nature of the lens galaxy might be a result of
an increased dark-matter fraction inside the Einstein radius from a
group halo. Groups have been found near a number of other lens
systems, hence they might not be uncommon and should be accounted for
in lens models where necessary (e.g.\ Lehar et al. 1997; Kundi\v{c} et
al. 1997a\&b; Tonry 1998; Tonry \& Kochanek 1999, 2000; Blandford,
Kundi\v{c} \& Surpi 2001; Keeton, Christlein \& Zabludoff 2000;
Fassnacht \& Lubin 2002; Johnston et al. 2003).

In our model, we associate the mass centroid with the light centroid
of the lens galaxy and model its mass distribution as an SIE with the
lens strength, position angle and ellipticity as free parameters. The
position angle is left free (as opposed to H1417) because the nearby
companions (Fig.~\ref{fig:HSTimages}) could introduce a difference
between the luminous and mass position angle. External shear is added.

\begin{figure*}[t]
\begin{center}
\leavevmode 
\vbox{
\hbox{
\epsfxsize=0.76\hsize 
\epsffile{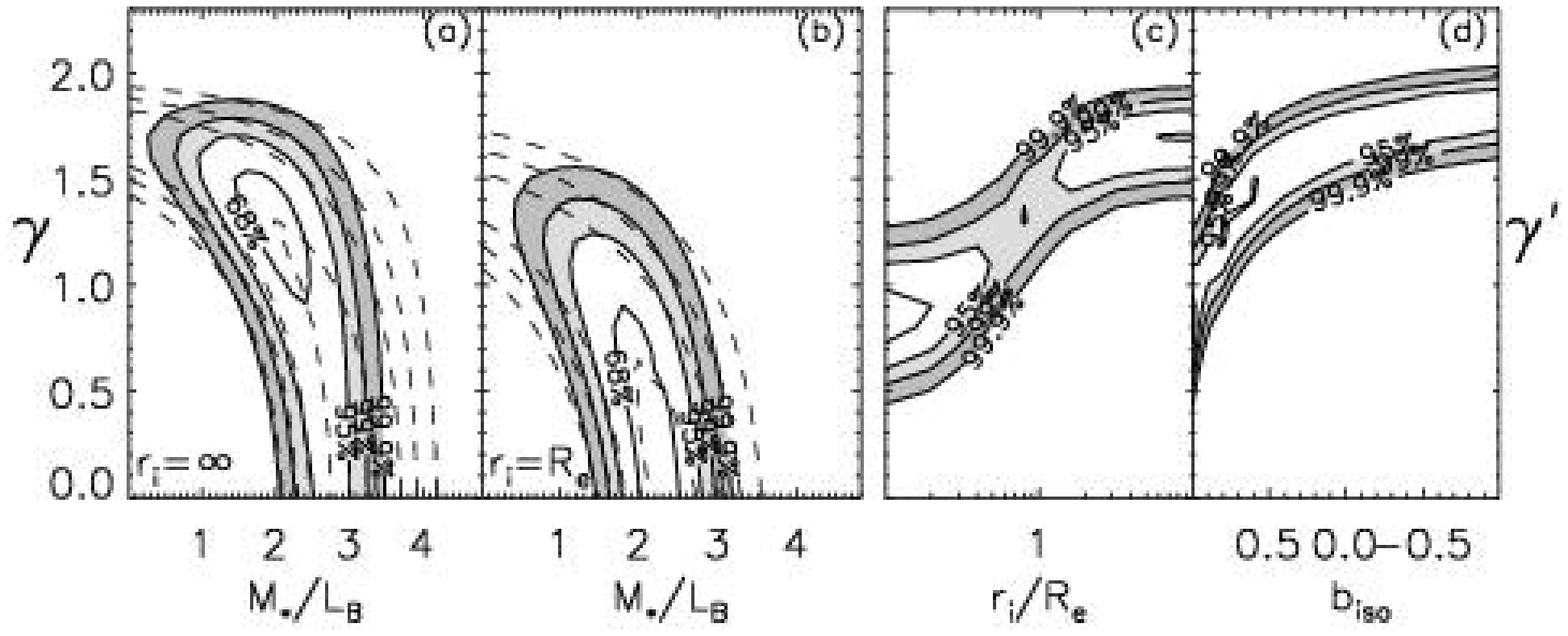}}
\hbox{
\epsfxsize=0.76\hsize 
\epsffile{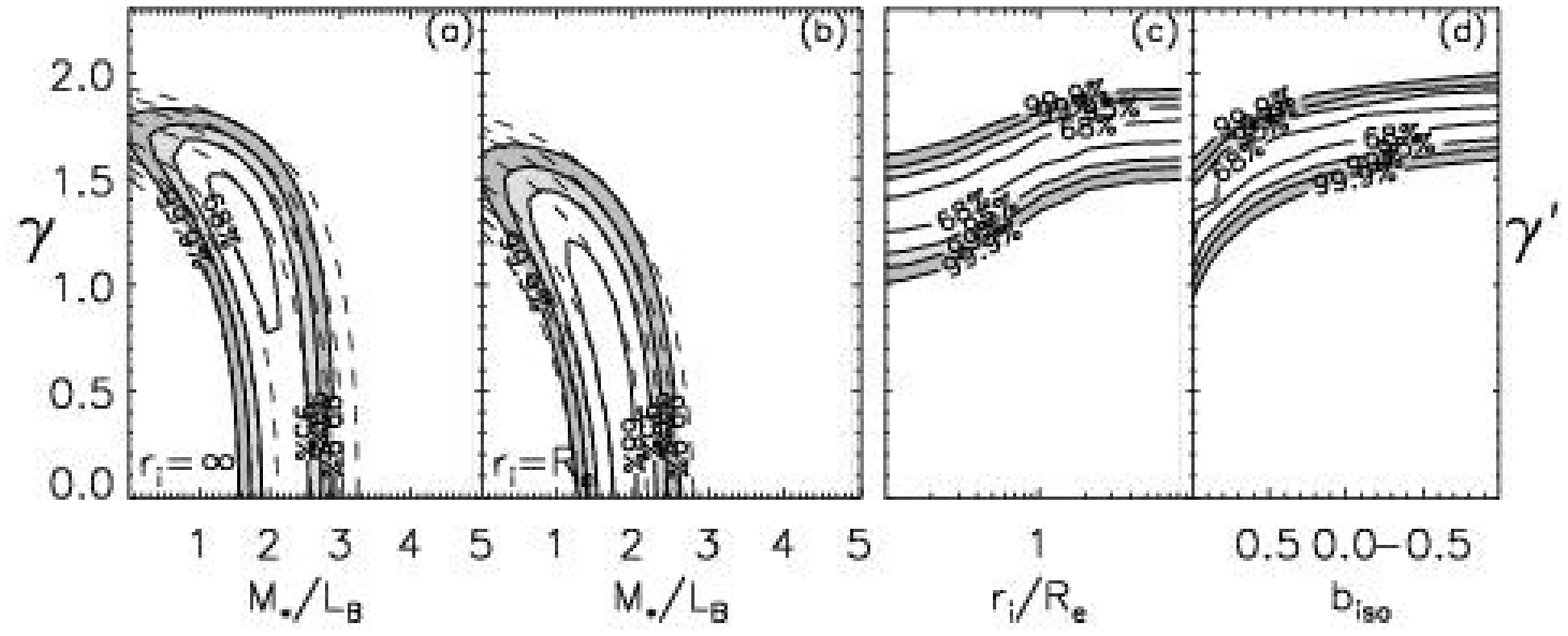}}
\hbox{
\epsfxsize=0.76\hsize 
\epsffile{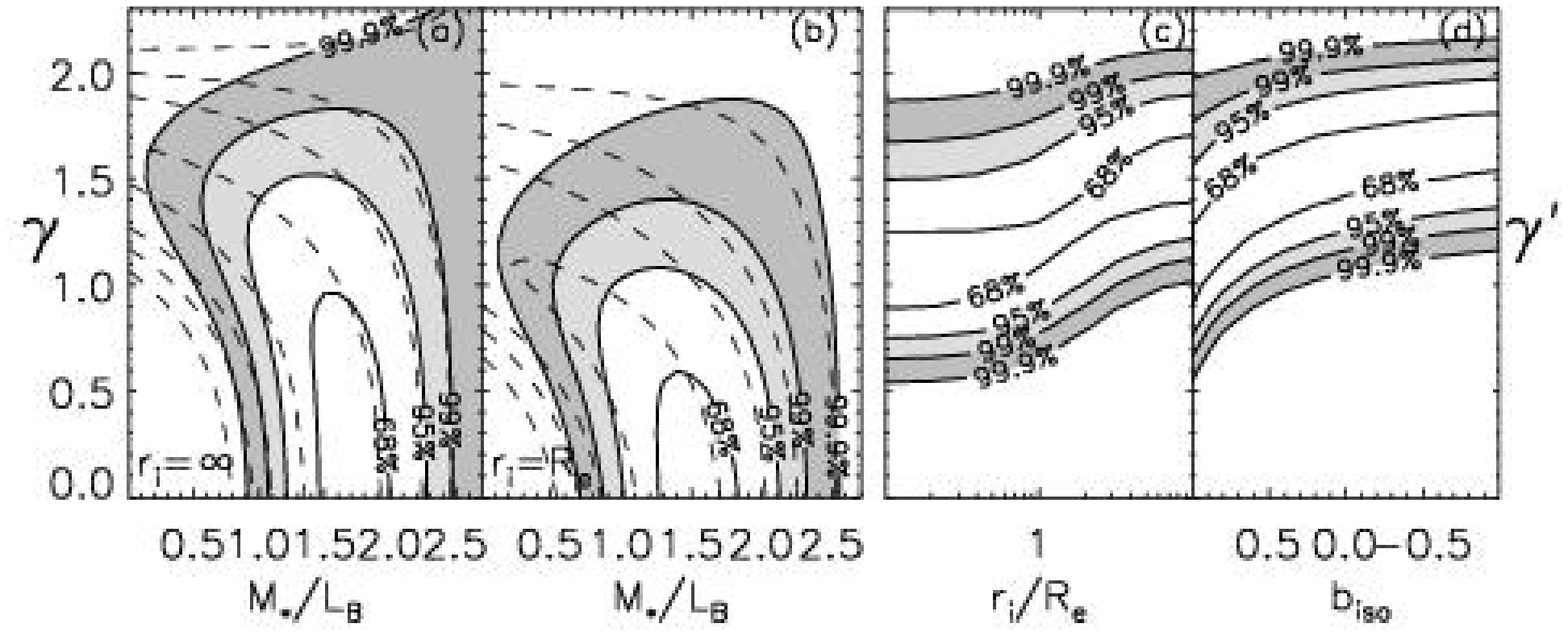}}
}
\end{center}
\figcaption{Likelihood contours for a joint lensing and dynamical
analysis of C0302 (upper row), H1417 (middle) and H1543 (lower). Panel
(a): The contours in the $M_{*}/L_{B}$-$\gamma$ plane for an isotropic
mass model, with (thick) and without (dashed) the FP constraints. Panel
(b): as (a) for a radially anisotropic model. Panel (c): contours in
the $r_i$ vs $\gamma'$ plane. Panel (d): contours in the
$\gamma'-b_{\rm iso}$ plane. Note that the effective slope $\gamma'$
changes very little for $b_{\rm
iso}=0\rightarrow-1$.\label{fig:chi0302}}
\end{figure*}

The source grid is $39\times$39 pixels (40~mas per pixel). Pixels in
the image plane within 1--$\sigma$ of the noise level are clipped and
do not participate in the determination of the best model (clipping at
2--3\,$\sigma$ results in a very similar model). Image-plane pixels
that map outside the source grid and visa versa are masked. We
minimize the reduced $\chi^2$, to properly account for the variable
number of participating source and image-grid pixels. We use an
appropriate PSF generated with Tiny-Tim. A value of $\lambda$=0.02 for
the smoothing parameter (see Appendix) leads to $\chi^2/{\rm
NDF}\approx1$.

Since the external shear is large, we find a strong degeneracy between
the lens-galaxy ellipticity and position angle and the external
shear. The constraint on $q=(b/a)$ is therefore relatively weak.
Since the external shear appears to align with the position angle of
the light distribution, not the mass model, we suspect that some of
the external shear is in fact `internal' shear due the stellar
component and that the lens galaxy is embedded in a larger misaligned
structure (e.g.\ a group halo). The presence of at least five small
dwarfs around C0320 would support this (Fig.~1). These degeneracies,
however, have negligible effect on the determination of enclosed mass
within the images.

The best SIE plus shear model is shown in the left panels of
Fig.\ref{fig:gridlens} and the mass-model parameters are listed in
Table~\ref{tab:lensmodels}.  We have refrained from calculating
precise formal errors -- as for H1417 -- which is extremely difficult
given the variability of NDF (i.e.\ the number of participating
pixels) and the free choice of source-pixel size and $\lambda$. It is
also computationally very expensive {\sl if} the mass models is not
fixed (see WD03 for the case when it {\sl is} fixed). We plan to
further refine the code -- currently written in IDL --, increase its
speed and allow for a full non-linear error analysis. Nevertheless, a
conservative upper limit of $\la$5\% can be set on the 1--$\sigma$
error on the Einstein radius, being roughly the width of the arc
divided by its distance to the lens centroid.  The equivalent SIS
velocity dispersion, mass and Einstein radius of the lens in C0302
from the best model then become, $\sigma_{\rm SIE}=294\pm8$~\kms,
$M_{\rm Einst}=(67.0\pm6.7)\cdot 10^{10}$~M$_\odot$ and $R_{\rm
Einst}=10.6\pm0.5$~kpc (1\farcs24$\pm$0\farcs06), respectively.

The dominant component of the source straddles the cusp on the ouside
and appears compact and relatively symmetric with possibly some
indication of extended structure around it.  The shape and position in
the source plane are similar to those found by Crampton et al.\
(2002), even though their stellar velocity dispersion, from a
non-singular model, is much higher than our SIE velocity
dispersion. This appears to be a result of the use of a non-zero core
radius and their definition of $\sigma$, which increases with
increasing core radius (Francois Hammer, private communication).  We
note that our definition of $\sigma_{\rm SIE}$ is in accordance with
previous work, including that of Kochanek et al. (2000) and Rusin et
al. (2003).

\subsection{HST15433+5352}

The modeling of H1543 proceeds very similar to that of C0302, with the
following differences.  First and foremost, there are several massive
nearby perturbers that have to be accounted for in the model. The
strongest is a massive galaxy (G2) approximately 4\farcs7 to the east
of the lens galaxy (G1) with a measured central stellar velocity
dispersion of $\sigma=263\pm11$~\kms. We model this galaxy as a SIS
(Eq.\ref{eq:kappa}\ with $q=1.0$) with $\sigma_{\rm SIE}=\sigma$.

Another galaxy (G3) at the same redshift of G1 and G2 was
serendipitously detected in the LRIS slit, 18$''$ away from G1 on the
opposite side of G2. Since H1543 falls near the edge of the WFPC2 field,
no HST images are available of G3. However, images of the region are
available from the Sloan Digital Sky Survey (SDSS) Data Release 1
(Abazajian et al. 2003). A visual inspection of the images shows
indeed a galaxy at the location of G3, and other galaxies with similar
colors and luminosities in the vicinity, consistent with the presence
of a group (see the discussion on groups for C0302).

Our best SIE plus shear model is shown in Fig.6. Relatively little
regularization is needed ($\lambda\sim0.003$). The source is
relatively compact, which lends further credit to the model. As
expected from the presence of the nearby aligned perturbers (G2 and
group), we find the external shear, $\gamma_{\rm ext}=0.17$, to be
large and dominant over the mass ellipticity of lens galaxy. To avoid
degeneracies between external shear and lens galaxy ellipticity, we
therefore restricted ourselves to a SIE mass model with $q=0.95$
fixed. We also tested SIE plus shear models with varying ellipticities
and find no strong differences between their critical and caustic
structures, nor between the enclosed mass within the equivalent SIS
Einstein radius.  The shear position angle ($\sim$72$^\circ$) aligns
nearly perfectly with the line between G1, G2 and the compact group
($\sim$67$^\circ$), suggesting the external shear is real and most
likely due to the group. The shear of G2 is already accounted for by
its SIS mass model.

\begin{inlinetable}
\begin{center}
\tabcaption{SIE gravitational lens models} \label{tab:lensmodels}
\begin{tabular}{lrrr}
  \hline
        & H1417 & C0302 & H1543\\
  \hline
  $x_{\ell}$ (arcsec) & $-$0.001$^{+0.004}_{-0.004}$  & - & -\\ 
  $y_{\ell}$ (arcsec) & $-$0.001$^{+0.004}_{-0.004}$ & - & -\\
  $b_{\ell}$ (arcsec) & 1.41$^{+0.10}_{-0.06}$ & 1.24 & 0.36 \\
  $q_{\ell}$ & 0.65$^{+0.05}_{-0.06}$ & 0.83 & $\equiv$0.95 \\
  $\theta_{\ell}$ ($^\circ$) & 31.7$^{+3.5}_{-4.0}$ & $-$33.4 & 16.0\\
  \hline
  $\gamma_{\rm ext}$ & 0.12$^{+0.01}_{-0.02}$ & 0.17 & 0.17 \\
  $\theta_{\rm ext}$ ($^\circ$) & 66.5$^{+3.4}_{-2.1}$ & $-$54.9 & 71.6 \\
  \hline
  $\nabla \kappa_{\rm g}$ (arcsec$^{-1}$) & 0.102$^{+0.015}_{-0.015}$ & - & -\\ 
  $\theta_{\rm g}$ ($^\circ$) & 200.4$^{+4.7}_{-5.7}$ & - & -\\
  \hline
\end{tabular}
\end{center}
{\footnotesize Note: The lens centers of C0302 and H1543 are fixed at
the observed galaxy centroids. The sky PA values are given.}
\end{inlinetable}

The equivalent SIS velocity dispersion, mass and Einstein radius of
the lens in H1543 from the best model are, $\sigma_{\rm
SIE}=139\pm7$~\kms, $M_{\rm Einst}=(3.4\pm0.7)\cdot 10^{10}$~M$_\odot$
and $R_{\rm Einst}=2.4\pm0.4$~kpc (0\farcs36$\pm$0\farcs04),
respectively. As for H1417 we can estimate the mass associated with
the external mass distribution from the external shear. If a mass
sheet with $\kappa_{\rm sheet}\sim \gamma_{\rm ext}\sim0.17$
contributed to the image separation, one would overestimate the mass
of the galaxy by $\sim$17\%. To account for this uncertainty, we adopt
a conservative error of 20\% on $M_{\rm Einst}$, twice that of the
other two systems. This range also covers the majority of models, when
varying different assumptions in the models (e.g. ellipticity).

We note that if G2 and the group are dynamically associated with the
lens galaxy, their dark-matter mass halos also contribute to the mass
inside the Einstein radius of the lens (G1), affecting both lensing
and stellar dynamics. Therefore, one should not regard this as a
systematic effect that should be removed like a mass-sheet, since this
mass truly contributes to the inner slope of the dark-matter halo of
the lens galaxy. We discuss this important point in more detail in
Sect.7.

\section{The two-component mass model}

\label{sec:mass}

Following our previous papers (TK02, KT03), in the lensing plus
dynamics analysis, we model the lens galaxies as a superposition of
two spherical components, one for the luminous stellar matter and one
for the dark-matter halo. The luminous mass distribution is described
by either a Hernquist (1990) 
\begin{equation}
  \rho_{\rm lum}(r)=\frac{M_*\,r_*}{2\pi\,r\,(r+r_*)^3}
\end{equation}
or a Jaffe (1983) model where $M_*$ is the total stellar mass. For
consistency with TK02 we will show primarily the results obtained with
the Hernquist profile and discuss how they change using a Jaffe
profile where relevant. The dark-matter halo is modeled as: ~\\
\begin{equation}
  \rho_{\rm DM}(r)=\frac{\rho_{\rm DM,0}\, r_{\rm
  b}^3}{r^{\gamma}\,(r^2_{\rm b}+r^2)^{(3-\gamma)/2}}
  \label{eq:DM}
\end{equation}
which closely resembles an NFW profile for $\gamma=1$, and has the
typical asymptotic behavior at large radii found from numerical
simulations of dark-matter halos $\propto r^{-3}$ (e.g. Ghigna et al.\
2000).  In accordance with the CDM picture (e.g. Bullock et al. 2001)
we expect the break radius $r_b$ to be much larger than the effective
and Einstein radii.  Therefore, in the following we will set $r_b\gg
R_{\rm Einst}$, effectively equivalent to $\infty$, i.e. we describe
the dark-matter halo as a simple power-law $\rho_{\rm DM}\propto
r^{-\gamma}$ in the region of interest. To further explore the effects
of the distribution of mass at large radii we have done tests with (i)
a dark-matter halo, as in Eq.~\ref{eq:DM}, but falling off as $r^{-4}$
at large radii (equivalent to a Hernquist or Jaffe model for
$\gamma=1,2$ respectively) and (ii) values of $r_b$ as small as $\la
R_{\rm e}$. In all cases we find that the effects on the stellar
velocity dispersion due to changes in the break radius and outer slope
are negligible if $r_b\ga 3\, R_{\rm e}$. In the current $\Lambda$CDM
models break radii as small as $r_b\la 3 R_{\rm e}$ are highly
unlikely for most galaxies, since $R_{\rm e}$ is typically only
several kpc. Our approximation of $r_b\rightarrow \infty$ is therefore
justified and the resulting constraints on $\gamma$ can be compared
with the inner dark-matter mass slope from simulations {\sl after}
baryonic collapse (e.g.\ cooling), which can alter the inner slope of
the dark-matter halo but much less so the outer slope.

We assume an Osipkov-Merritt (Osipkov 1979; Merritt 1985a,b)
parametrization of the anisotropy of the stellar mass distribution or
a constant $\beta(r)$ model
\begin{equation}
 \beta(r)=\left\{
\begin{array}{ll}
  1-\frac{\sigma^2_{\theta}}{\sigma_{r}^2}=\frac{r^2}{r^2+r^2_i} & r_i\ge 0\\
  ~\\
  b_{\rm iso}\in [-1,+1] & \\
\end{array} \right.
\label{eq:OM}
\end{equation}
where $\sigma_{\theta}$ and $\sigma_{r}$ are the tangential and radial
components of the velocity dispersion and $r_i$ is called the
anisotropy radius. Note that $\beta\ge0$ by definition for the OM
model, not allowing for tangentially anisotropic models. At infinite
radii, Osipkov-Merrit models become completely radial. Although this
behavior is not commonly found within the inner regions of E/S0
galaxies probed by observations (e.~g. Gerhard et al.\ 2001; see also
van Albada 1982 and Bertin \& Stiavelli 1993 for theoretical grounds),
it has little impact in the case considered here, since the pressure
tensor only becomes significantly radial well outside the Einstein
Radius and in projection is significantly down-weighted by the rapidly
falling luminosity--density profile. To test tangential anisotropy, we
also considered models with constant anisotropy $\beta(r)=b_{\rm iso}$
varying from $-$1 to $+$1. Whereas for $b_{\rm iso}=1 \rightarrow 0$
the behavior is very similar to $r_i=0 \rightarrow \infty$, the effect
on the infered mass slope $\gamma'$ for $\beta=-1$ to 0 is almost
negligible (panels (d) in Fig.7). We will therefore not further
consider these models in detail, but only mention them when necessary.

\begin{inlinefigure}
\begin{center}
\resizebox{\textwidth}{!}{\includegraphics{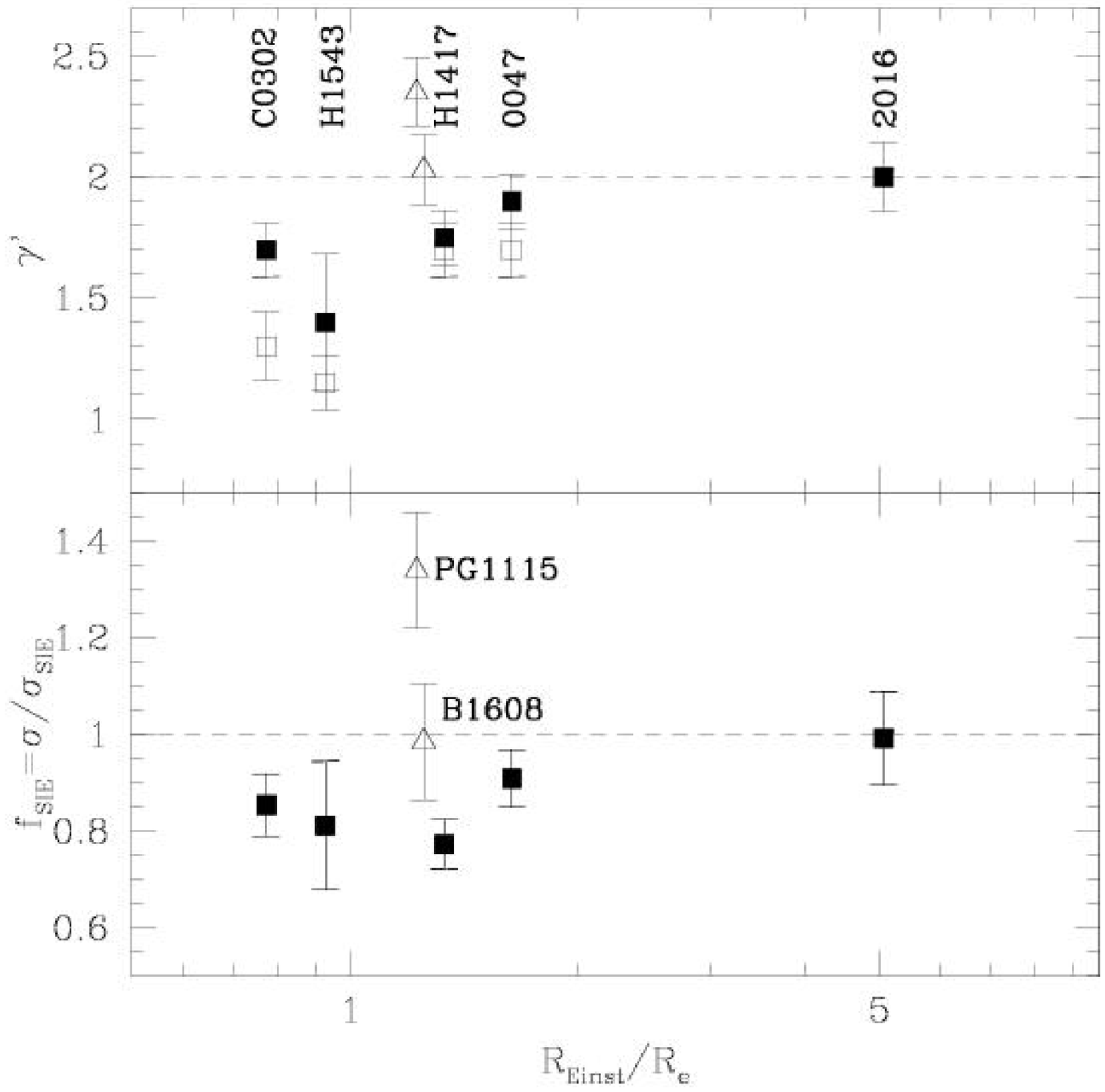}}
\end{center}
\figcaption{Upper panel: The effective slope as a function of the
ratio between the Einstein Radius and the effective radius; solid
points are obtained for isotropic models ($r_i=\infty$), open points
for radially anisotropic models ($r_i=R_e$). Lower panel: The ratio
between central velocity dispersion ($\sigma$) and velocity dispersion
of the best fitting SIE model as a function of the ratio between the
Einstein Radius and the effective radius. The horizontal dashed lines
represent the expected values for isothermal lens models. The measured
values (isotropic models only) for PG1115 (TK02b) and B1608 (K03) --
not included in the LSD sample -- are also shown for completeness as
open triangles. See Sections 6 and 7 for discussion.
\label{fig:aplotap}}
\end{inlinefigure}

The line-of-sight velocity dispersion is obtained by solving the three
dimensional spherical Jeans equation for the luminous component in the
total gravitational potential and computing the luminosity-weighted
average along the line of sight (e.g.\ Binney \& Tremaine 1987;
Kochanek 1994). We correct for the average seeing during the
observations, and average the velocity dispersion -- weighted by the
surface brightness -- inside the appropriate rectangular
apertures. For completeness, we rescale the model apertures such that
their projection on the axisymmetric model is equivalent to their
projection on an elliptical galaxy with an axial ratio of $b/a$. This
has minimal effects on the model velocity dispersions -- much smaller
(i.e.\ $<1$\%) than the observational errors -- since the observed and
model dispersion profiles are typically very flat. The uncertainties
on seeing, aperture size, and galaxy centering are taken into account
as systematic errors in the following discussion. Additional
discussion of our mass profiles and dynamical model can be found in
TK02 and KT03.

\section{A joint lensing and dynamical analysis} 
\label{sec:analysis}

We are now in the position to use the measurements derived in
Sections~\ref{ssec:photo},~\ref{ssec:spec} and~\ref{sec:lens} to
constrain the free parameters in our two component mass
models. Surface photometry gives directly $r_*=R_{\rm e}/1.8153$ for
the Hernquist model, assuming that the stellar mass-to-light ratio is
constant. The mass enclosed by the Einstein radius is used to obtain
$\rho_{\rm DM,0}$, given the other parameters. Likelihood contours of
the three remaining parameters (i.e.\ $M_*/L_B$, $\gamma$ and $r_i$)
are then obtained by comparing the velocity dispersion profiles from
the models with the observed ones.

\subsection{The FP as an additional constraint}
\label{ssec:FPadd}

An additional constraint can be obtained using the offset of the
galaxy from the local FP by introducing one further assumption.  If
the evolution of the effective mass to light ratio $\Delta \log
(M/L_B)$ is equal to the evolution of the stellar mass-to-light ratio
$\Delta \log (M_*/L_B)$, the stellar mass-to-light at redshift $z$ is
related to the stellar mass-to-light ratio at $z=0$ by

\begin{equation}
     \log \left(\frac{M_*}{L_B}\right)_{z}= \log
     \left(\frac{M_*}{L_B}\right)_{0} + \Delta \log
     \left(\frac{M}{L_B}\right),
\label{eq:FPev}
\end{equation}

\noindent where the first term on the right hand side of the equation
can be measured for local E/S0 galaxies. Using the local value of
$(M_*/L_B)_0=(7.3\pm2.1)\,h_{65}$~\mlu, determined from data by
Gerhard et al.\ (2001) as in TK02, we infer $M_*/L_B=(1.9\pm0.5)\,
h_{65}$~\mlu, $M_*/L_B=(1.6\pm0.4)\,h_{65}$~\mlu,
$M_*/L_B=(2.1\pm0.6)\,h_{65}$~\mlu\ for the lens galaxies in C0302,
H1417 and H1543, respectively. In the next sections we will compare
these measurements with the {\it independent} ones obtained from the
joint lensing and dynamics analysis, finding a good agreement. Then,
we will also use the FP measurements as a further constrain to the
two-component models. However -- since this determination relies on a
non-trivial assumption (Eq.~\ref{eq:FPev}) -- we will present both the
results that include this constraint and those that do not.

\begin{table*}
\begin{center}
\caption{Summary of lens/dynamical model results -- I \label{tab:lens2}}
\begin{tabular}{lccccccc}
\hline
\hline
Galaxy & $\gamma'$ (iso)        & $\gamma'$ (aniso)      & $\sigma_{\rm SIE}$  & $f_{\rm SIE}$ & R$_{\rm Einst}$ & $R_{\rm e}$ \\
       &                        &                        & (\kms)              &               & (arcsec)        & (arcsec)    \\
\hline
0047   &  1.90$\pm$0.05$\pm$0.1 & 1.7$\pm$0.05$\pm$0.10  &  252$\pm$4          & 0.91$\pm$0.06  & 1.34$\pm$0.01  &  0.82$\pm$0.12 \\
C0302  &  1.70$\pm$0.05$\pm$0.1 & 1.3$\pm$0.10$\pm$0.10  &  294$\pm$8          & 0.85$\pm$0.06  & 1.24$\pm$0.06  &  1.60$\pm$0.15 \\
H1417  &  1.75$\pm$0.05$\pm$0.1 & 1.7$\pm$0.05$\pm$0.10  &  290$\pm$8          & 0.77$\pm$0.05  & 1.41$\pm$0.08  &  1.06$\pm$0.08 \\
H1543  &  1.40$\pm$0.20$\pm$0.2 & 1.15$\pm$0.05$\pm$0.10 &  139$\pm$7          & 0.83$\pm$0.13  & 0.36$\pm$0.04  &  0.41$\pm$0.04 \\
MG2016 &  2.00$\pm$0.10$\pm$0.1 & 2.0$\pm$0.10$\pm$0.10  &  331$\pm$10         & 0.99$\pm$0.10  & 1.56$\pm$0.02  &  0.31$\pm$0.06 \\
\hline 
\end{tabular}
\end{center}
\end{table*}

\subsection{Single power-law mass models}
\label{ssec:power}

Before considering the full two-component models (Section.~5) let us
first consider a simplified family of models to explore the properties
of the total mass distribution, since this is of particular relevance
to studies of, for example, the value of H$_0$ from time-delays and
also lensing statistics.

As in TK02 and KT03 this family of models consists of a total luminous
plus dark-matter mass distribution that follows a single power-law
$\rho_{\rm tot}\propto r^{-\gamma'}$ within the region of interest,
where $\gamma'$ is called the {\sl effective slope}. Hence, the
luminous mass is assumed to be a trace component in the potential,
with $M_*/L_{\rm B}=0$. The two remaining free parameters, $\gamma'$
and $r_i$ are constrained with the velocity dispersion profile,
yielding the results shown in panels (c) of Fig.~\ref{fig:chi0302}.
In panel (d), the constant $\beta$ models are shown, displaying a
similar behavior for $b_{\rm iso}\ge0$, whereas for $b_{\rm iso}<0$
(tangential anisotropy) almost no effect is seen on the value of
$\gamma'$.

The best-fit values of $\gamma'$ depend on the anisotropy of the
velocity ellipsoid. As expected, an isotropic velocity ellipsoid
($r_i=\infty$ or $b_{\rm iso}=0$) leads to a larger value of
$\gamma'$, whereas for the more radial orbital structures ($r_i
\rightarrow 0$), a smaller value of $\gamma'$ is needed. The most likely
values of $\gamma'$ for two representative cases ($r_i=\infty$ and
$r_i=R_{\rm e}$) are listed in Table~\ref{tab:lens2}. The
corresponding values for MG2016 and 0047 (TK02a, KT03) are also listed
for completeness.  

The results are also shown in Figure~\ref{fig:aplotap}, where we plot
$\gamma'$ as a function of $R_{\rm Einst} / R_{\rm e}$. The average
slope from the five systems in our sample is $\langle\gamma'\rangle
=1.75\pm0.09$ with a large rms of 0.20 (isotropic), or $\langle
\gamma'\rangle =1.57\pm0.16$ with an rms of 0.35
(anisotropic). Extremely radial orbits ($r_i\la R_{\rm e}$) can
probably be ruled out, both on observational grounds (e.g.\ Gerhard et
al.\ 2001) and on theoretical grounds, since they would lead to
instabilities (Merritt \& Aguilar 1985; Stiavelli \& Sparke 1991),
whereas tangential anisotropy can not be ruled out, it has a
negligible effect (see Panels (d) in Figure~7).  

In the lower panel of Figure~\ref{fig:aplotap}, we also show $f_{\rm
SIE}$, i.e.\ the ratio between the central velocity dispersion and the
velocity dispersion of the best fitting Singular Isothermal Ellipsoid.
This number is independent of the choice for the dynamical model.  The
average is $\langle f_{\rm SIE}\rangle =0.87\pm0.04$ with an rms of
0.08, lower on average than that based on the expectation that
$\sigma\approx \sigma_{\rm SIE}$ (Kochanek 1994; Kochanek et al.\
2000).

\subsection{Luminous and dark-matter mass decomposition}
 
Let us now consider the two component mass models. Once again, we
examine the two cases of $r_i=\infty$ and $r_i=R_{\rm e}$, delineating
a conservative range of physical models, and derive likelihood
contours in the $M_*/L_B$--$\gamma$ plane. The likelihood contours are
shown in the panels (a) and (b) of Fig.~\ref{fig:chi0302}. The dashed
lines represent the contours obtained without the FP constraint on
$M_*/L_B$, while the solid contours include the FP constraint
(Sect.6). The main effect of including the FP constraint is to rule
out regions with low stellar mass-to-light ratios. In general, the
shape of the contours is well understood: the outer luminous component
is on average steeper than the dark-matter component and therefore
smaller values of $M_*/L_B$ require larger values of $\gamma$, to
compensate and produce a total mass profile as steep as required by
the kinematic data. Increasing the radial anisotropy implies a
more shallow total mass profile and therefore a smaller value of
$\gamma$.

In this subsection we discuss constraints on the amount of dark matter
within the Einstein radius, through measurements of the total and
stellar mass to light ratio. In the next subsection
(Sect.~\ref{ssec:MLev}), we will use these values to determine the
cosmic evolution of the stellar mass to light ratio and hence the star
formation history of E/S0 galaxies.  In light of these two goals we
will determine the fraction of dark matter marginalizing the
likelihood contours shown in Figure~\ref{fig:chi0302} over $\gamma$ in
two ways: (i) Including the FP constraint as a prior, to obtain the
most precise measurement of $M_*/L_B$. (ii) Not including the FP
constraint, to obtain an independent measurement of the cosmic
evolution of $M_*/L_B$.  The results are listed in
Table~\ref{tab:lens1}. 
The 68\% confidence limits -- around the
maximum-likelihood value -- are uniquely determined from the
probability distribution function of $M_*/L_B$ (after marginalising
over $\gamma$) by the two values of $M_*/L_B$ that have {\sl equal}
probability densities and enclose 0.68 of the probability. Since for
low values of $M_*/L_B$, the probability density as function of
$\gamma$ is only large for a small range around $\gamma \sim 2$, it is
nearly constant over a large range of $\gamma$ for $M_*/L_B\sim 2-3$
(Fig.~\ref{fig:chi0302}). Hence, after marginalisation, $M_*/L_B$ has
a clearly-defined maximum-likelihood value and 68\% confidence
limits.

The total mass to light ratio enclosed within the Einstein radius
$(M_{\rm tot}/L_B)(<R_{\rm Einst})$ is also listed in
Table~\ref{tab:lens1} for comparison. Note that $(M_{\rm
tot}/L_B)(<R_{\rm Einst})$ is considerably higher than the limit on
$M_*/L_B$ for all five E/S0 galaxies, implying that the galaxies
cannot be described by a constant mass-to-light ratio model. Hence,
the presence of a mass component spatially more extended than the
luminous component is required. We identify this component with the
dark-matter halo.  Quantitatively, we find that for all five E/S0
galaxies (including MG2016 and 0047), the {\sl no-dark-matter-halo}
scenario is excluded at the $>99\%$ C.L. 
In other words, a
constant mass-to-light ratio model is too steep to satisfy
simultaneously the lensing and dynamical constraints and can therefore
be ruled out at the 99\% C.L.

Finally, we note here that $M_*/L_B$ and $M_{\rm tot}/L_B(<R_{\rm
Einst})$ are correlated through $L_B$, which is only a scaling factor,
useful for the physical interpretation, but irrelevant for the
lensing+dynamical analysis. Thus it is preferrable to express our
results in terms of the fraction of dark matter $f_{\rm
DM}=(1-M_*/M_{\rm tot})$, or equivalently in terms of the fraction of
luminous matter $f_*=1-f_{\rm DM}$. In the isotropic case, the range
of dark-matter mass fractions inside the Einstein radius is $f_{\rm
DM}(<R_{\rm Einst})$=0.37--0.72. Translating this into a mass fraction
inside $R_{\rm e}$ is slightly model-dependent. However for
$\gamma$=0--1, the range of dark-matter mass fractions is $f_{\rm
DM}(<R_{\rm e})$=0.15--0.65, with at most a 10\% change (both ways) in
the value of $f_{\rm DM}$ between $\gamma=1$ and $\gamma=0$. This
confirms our conclusion that all five E/S0 lens galaxies at $z\approx
0.5-1$ have massive dark-matter halos, even well inside the luminous
component.  Note that these values for the dark-matter fraction are
significantly higher than the limits obtained from lensing statistics
for adiabatically contracted lenses, $f_{\rm DM}(<R_{\rm e})<33\%$
(95\% CL; Keeton 2001).

\begin{table*}
\begin{center}
\caption{Summary of lens/dynamical model results. II.\label{tab:lens1}}
\begin{tabular}{lccccccc}
\hline \hline Galaxy &  M$_{\rm E}$ & (M$_{\rm tot}$/L$_{\rm
B}$)$_{<R_{\rm E}}$ & (M$_*$/L$_{\rm
B}$)$^{\rm ld+FP}_{\rm iso}$ & (M$_{*}$/L$_{\rm B}$)$_{\rm iso}^{\rm ld}$ &
$f_{\rm DM}^{\rm ld+FP}$ (iso) &
$f_{\rm DM}^{\rm ld}$ (iso) \\ 
& (10$^{10}$M$_{\odot})$ & \mlu & \mlu & \mlu & ($<R_{\rm Einst}$) & ($<R_{\rm Einst}$)\\ 
\hline 
0047  &  40.6$\pm$2.0   & 5.4$\pm$0.5 & 3.0$^{+0.3}_{-0.6}$ & 3.0 $^{+0.4 }_{-1.1}$ & 0.44$^{+0.11}_{-0.14}$ & 0.44$^{+0.12}_{-0.22}$ & \\ 
C0302 &  67.0$\pm$6.7   & 4.8$\pm$0.5 & 2.2$^{+0.5}_{-0.5}$ & 2.8 $^{+0.7 }_{-0.8}$ & 0.54$^{+0.15}_{-0.15}$ & 0.42$^{+0.18}_{-0.20}$ & \\ 
H1417 &  70.6$\pm$7.0   & 5.0$\pm$0.5 & 1.9$^{+0.1}_{-0.2}$ & 2.1 $^{+0.3 }_{-0.2}$ & 0.62$^{+0.10}_{-0.11}$ & 0.58$^{+0.12}_{-0.11}$ & \\ 
H1543 &  3.4$\pm$0.7    & 2.7$\pm$0.5 & 1.7$^{+0.3}_{-0.4}$ & 1.5 $^{+0.4 }_{-0.7}$ & 0.37$^{+0.22}_{-0.24}$ & 0.44$^{+0.24}_{-0.32}$ & \\ 
2016  &  110.0$\pm$11.0 & 8.0$\pm$0.8 & 2.2$^{+0.3}_{-0.3}$ & 2.5 $^{+0.3 }_{-0.4}$ & 0.72$^{+0.11}_{-0.10}$ & 0.69$^{+0.11}_{-0.11}$ & \\ 
\hline
\end{tabular}
\end{center}
{\footnotesize Notes: The mass-to-light ratio is marginalized over
$\gamma>$0.}
\end{table*}

\subsection{The evolution of $M_*/L_B$ from lensing \& dynamics}

\label{ssec:MLev}

The sample of E/S0 lens galaxies reaches a large enough redshift to
afford a direct measurement of the evolution of the stellar
populations of E/S0 without including any constraint from the FP. In
Figure~\ref{fig:MLev}, we have plotted $M_*/L_{\rm B}$
(Table.~\ref{tab:lens1}) as a function of redshift. The stellar
mass-to-light ratios of E/S0 galaxies at $z$\,$\approx$\,0.5--1.0 are
significantly smaller than in the local Universe (on average
$2.3\pm0.6$ \mlu\ versus $7.3\pm2.1$ \mlu) implying considerable
ageing of the stellar populations in the last 4--8 Gyrs.

\begin{inlinefigure}
\begin{center}
\resizebox{\textwidth}{!}{\includegraphics{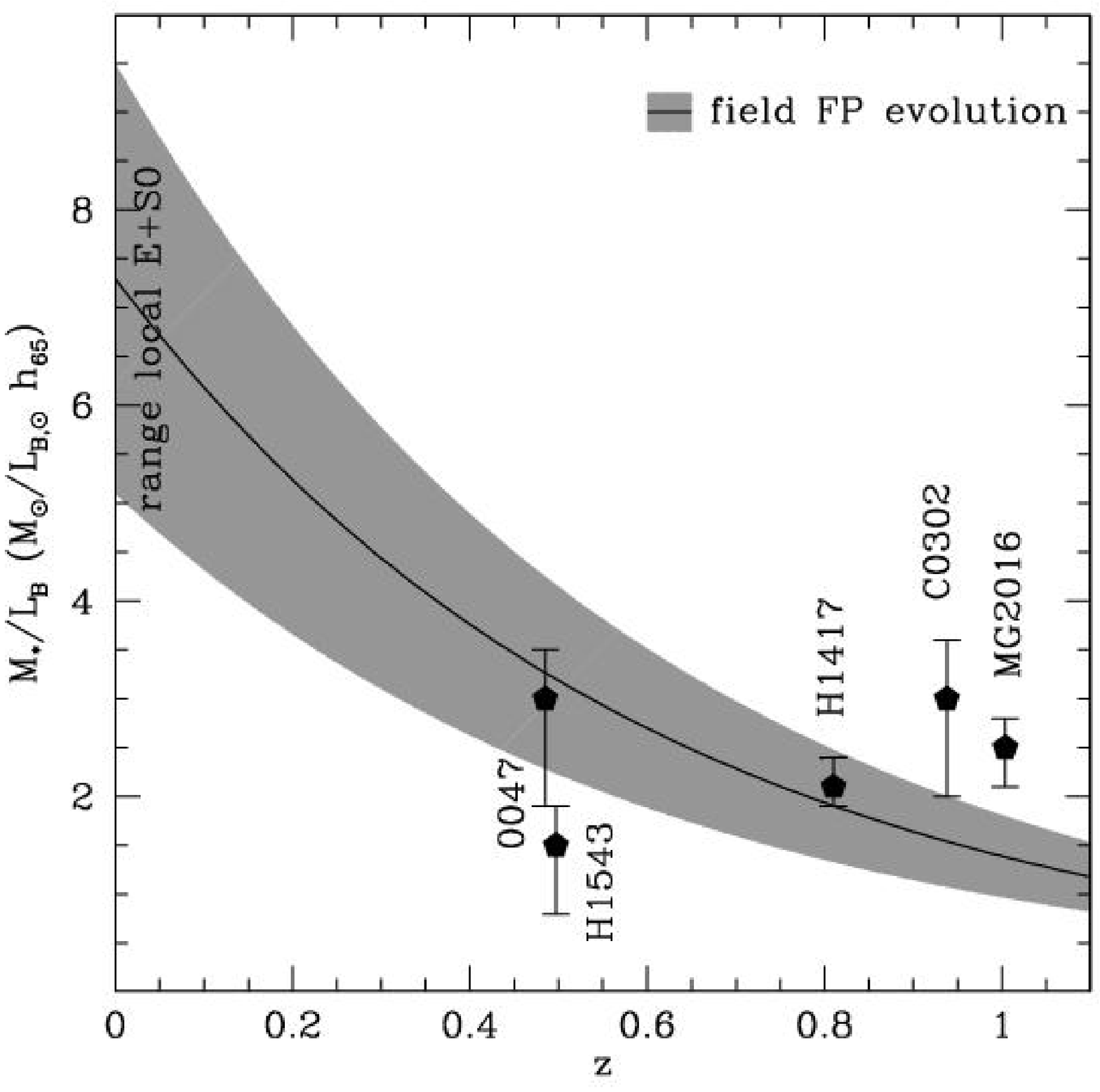}}
\end{center}
\figcaption{The cosmic evolution of the stellar mass-to-light ratio
(for an isotropic velocity ellipsoid, $r_i=\infty$). The solid points
are the results from the joint lensing \& dynamics analysis, whereas
the shaded region is the independent measurement via the Fundamental
Plane from Treu et al. (2002). We emphasize that the latter is not
based on lens galaxies and the former not on the FP measurements of
$M_*/L_B$ from the five lenses as shown in Fig.\ref{fig:FP}.
\label{fig:MLev}}
\end{inlinefigure}

In Fig.~\ref{fig:MLev}, we also compare our direct lensing+dynamics
measurement of the stellar mass-to-light ratio (Table~\ref{tab:lens1})
with the indirect measurement obtained from the evolution of the FP of
field E/S0 (Treu et al.\ 2002). The agreement is very good, consistent
with a scenario of pure luminosity evolution of E/S0 from $z\sim1$ to
today. This measurement rules out scenarios predicting strong
evolution of the internal structure of E/S0 galaxies with redshift,
where the virial coefficient relating $\sigma^2 R_{\rm e}$ to the
stellar mass would change significantly with cosmic time.

If we use the value of $(M_*/L_{\rm B})_{\rm iso}^{\rm ld}$ from
Table~\ref{tab:lens1} to determine the evolution of the stellar
mass-to-light ratio with redshift, we find that the average evolution
is $\langle d \log ( M/L_{\rm B})/d z\rangle =-0.75\pm0.17$, in good
agreement with the results from Sect.~\ref{sec:mlevfp}, which gave $d
\log( M/L_{\rm B})/ d z = -0.72\pm0.10$ based on the FP. A
disagreement between the two independent results would have implied
that either the FP is not a good method to derive $M_*/L_B$ evolution
or that our lensing+dynamical analysis is faulty.

\subsection{The inner slope of the dark-matter halos}

Since the values of $M_*/L_B$ agree between determinations from the FP
and lensing plus dynamics, we can feel confident that their
measurements can be combined, as shown by the solid lines in panels
(a) and (b) in Fig.\ref{fig:chi0302}. The posterior probability
distributions function of $\gamma$ -- marginalized over \ml, including
the FP constraint -- for the five lenses are shown as solid colored
lines in Figure~\ref{fig:gammap}.

For individual lenses these posterior probability distribution
functions imply upper limits on $\gamma$ between $\sim$1 and
$\sim$1.5, i.e.\ consistent with the inner cusps predicted by
cosmological simulations, if the collapse of baryons to form stars did
not significantly steepen the dark-matter halo (see TK02, KT02; see
also Loeb \& Peebles 2003; Sand et al.\ 2002, 2004; El-Zant et
al. 2003; Nipoti et al.\ 2004). Somewhat tighter confidence limits are
obtained when combining the measurements (dashed line):
$0.97<\gamma<1.46$ or most likely $\gamma = 1.3^{+0.2}_{-0.4}$
(isotropic) $0<\gamma<0.62$ (anisotropic) at 68\%CL and are
$0.39<\gamma<1.59$ (isotropic) $0<\gamma<1.26$ (anisotropic) at 95\%
CL. In conclusion, the slope of the dark-matter halos is definitely
flatter than isothermal, and ranges between the value predicted by
numerical simulations and zero, depending on anisotropy. Requiring
consistency with numerical simulations implies that (i) significant
radially anisotropic models (i.e.\ $r_i\approx R_e$) are ruled out and
(ii) dark-matter halos do not significantly steepen during baryonic
collapse. More stringent statements cannot be made at this stage,
because uncertainties related to the orbital properties of the stars
dominate the error budget\footnote{An additional source of uncertainty
is due to the mass profile of the luminous component: the contours of
$\gamma'$ shift towards slightly lower values by adopting a Jaffe
model for the luminous component.}.

\section{The homogeneity of early-type galaxies}

\label{sec:homo}

Now that we have homogeneously analyzed a sample of five lenses at
$z\approx 0.5-1.0$, we can start to look into the general properties
of the E/S0 lens galaxies\footnote{We note that the lens galaxies are
not a statistical sample by any means, so the results should be
interpreted as an exploration of the variety of possible behaviors
rather than in a rigorous statistical sense.}.

\subsection{Are E/S0 galaxies isothermal?}

The first important question that we want to discuss is the average
total mass density profile of lens galaxies, i.e. what is the
distribution function of $\gamma'$ (see
Section~\ref{ssec:power}). This is relevant not only in terms of
formation scenarios, but also in many application of lensing. For
example, cosmological parameters from lens statistics are generally
obtained assuming that lenses are isothermal (e.g.\ Turner et
al. 1984; Fukugita et al. 1990; Kochanek 1996; Helbig et al. 1999;
Falco et al. 1999; Chae et al.\ 2002; Mitchell et
al. 2004). Similarly, the Hubble Constant from gravitational
time-delays (Refsdal 1964) is typically obtained assuming isothermal
mass density profile as well (e.g.\ Kundi\v{c} et al. 1997a; Schechter et
al. 1997; Impey et al. 1998; Biggs et al.\ 1999; Koopmans \& Fassnacht
1999; Koopmans 2001; Kochanek 2002; Wucknitz et al.\ 2003). Also
estimating the central stellar velocity dispersion $\sigma$ from lens
models depends on the assumed mass model. Hence, how justified is the
isothermal approximation?

\begin{inlinefigure}
\begin{center}
\resizebox{\textwidth}{!}{\includegraphics{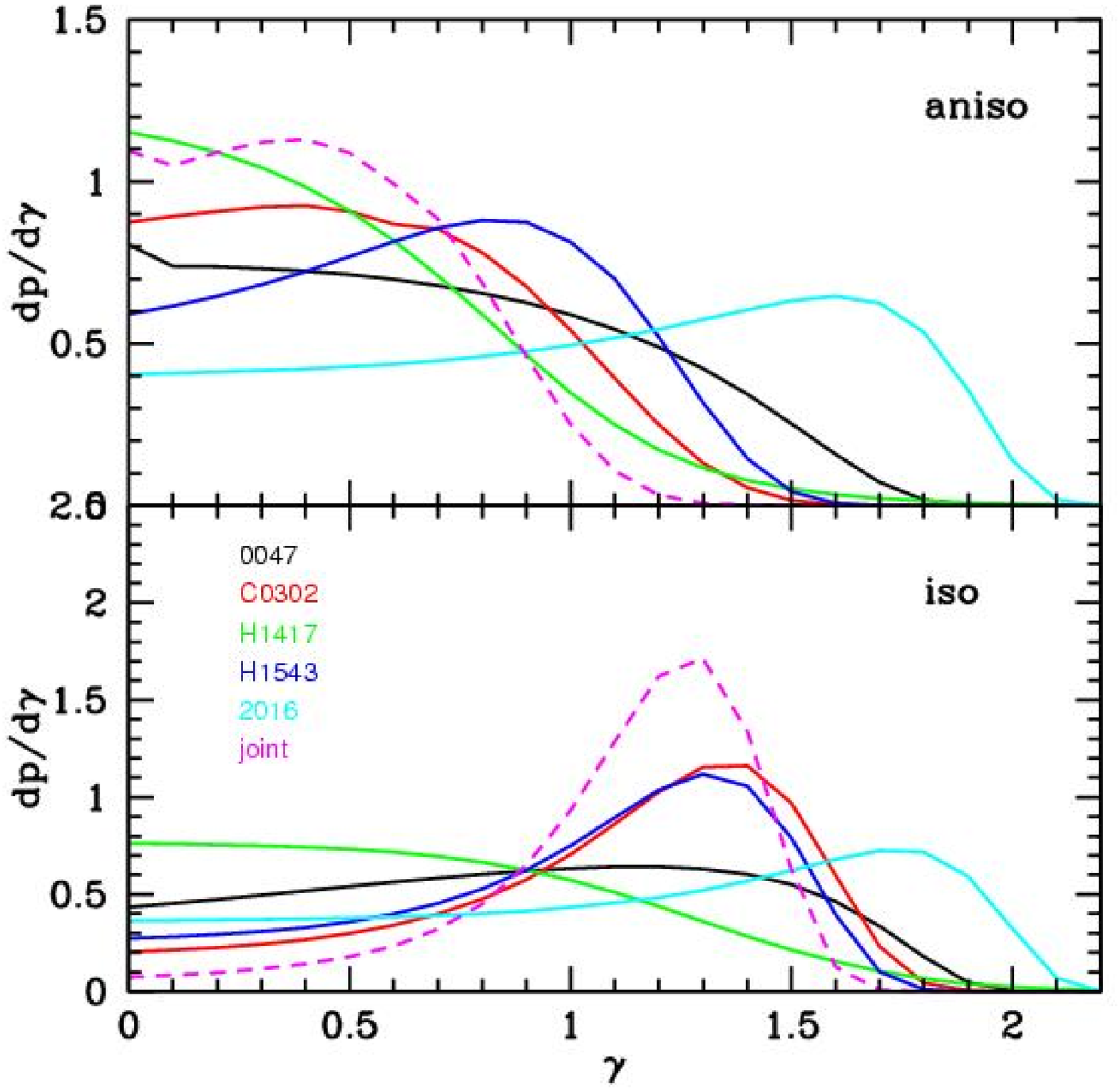}}
\end{center}
\figcaption{\label{fig:gammap} Posterior distribution functions for
$\gamma$ for the individual lenses and the joint probability. The
case for isotropic orbits is shown in the lower panel, the case for
radially anisotropic orbits is shown in the upper panel.}
\end{inlinefigure}

The main conclusion from Section~\ref{ssec:power} is that there
appears to be intrinsic scatter in the values of $\gamma'$ for lower
values of $R_{\rm Einst}/R_e$. A possible trend is seen, with
$\gamma'$ increasing with $R_{\rm Einst}/R_e$ and reaching
$\gamma'\approx2$ at large relative radii. However, based only on five
systems, it is dangerous to interpret this as a real physical
trend. For example, the two other lenses for which we have done a
similar analysis PG1115+080 (TK02b) and B1608+656 (K03)-- not selected
as part of the LSD Survey\footnote{Because they fail to meet the
criteria of favorable contrast between the lens and the source.} --
have $\gamma'>2$ ($\gamma'=2.35\pm0.1\pm0.1$ and
$\gamma'=2.03\pm0.1\pm0.1$ respectively). By including those in the
plot (open triangles in Figure~\ref{fig:aplotap}), the scatter in
$\gamma'$ would increase at low values of $R_{\rm Einst}/R_e$, erasing
any apparent trend. The straight average for this extended sample
becomes $\langle \gamma'\rangle=1.9 \pm 0.1$ with an rms of 0.3 for
the isotropic case ($r_i=\infty$). Hence, with a larger sample in
hand, one might indeed find that $\gamma'$ approaches $\approx 2$, but
also that there is significant intrinsic scatter (see also Rusin et
al. 2003, who statistically build a mass profile from a sample of lens
galaxies and find $\langle \gamma'\rangle=2.07\pm0.13$). Is such a
large scatter expected?  At the radii of interests, typically of the
order of the effective radius, we can expect scatter for at least two
reasons.

First, E/S0 galaxies are typically found in or near groups or clusters
and we expect that E/S0 lens galaxies will be in a similar environment
(see discussion on the C0302 lens model). A group dark-matter halo
might be present around the lens (a cluster will be external, since
they are critical themselves and would produce much larger image
separations). If the inner mass slope of the group halo is shallower
than isothermal (as typically found in the inner regions of
groups/clusters; e.g. Ettori et al. 2002; Kelson et al. 2002; Gavazzi
et al. 2003; Kneib et al. 2003; Sand et al. 2002, 2004), it could
introduce a `floor' of dark matter that will result in $\gamma'<2$
(corresponding to the rising velocity dispersion profiles observed in
some elliptical galaxies at the center of clusters, e.g. Dressler
1979; Carter et al. 1985).  Arguments suggesting that H1543, C0302 and
possibly H1417 might be located in or near groups were presented in
the previous sections.  We emphasize that this is different from an
intervening mass-sheet, since the group is physically centered on or
close to the lens galaxy and thus also affects the stellar dynamics.
On the other hand, if the Einstein radius is small enough, the total
mass distribution will become more dominated by the luminous
component, which is typically steeper than isothermal and results in
$\gamma'>2$. At face value, the trend seen in Figure~\ref{fig:aplotap}
is opposite to what we would expect: for small $R_{\rm Einst}/R_e$,
where baryon dominate we would expect $\gamma'>2$, while at larger
radii, where the group halo dominates, we would expect $\gamma'<2$.
Given the present size of the sample of five systems, the argument is
inconclusive. The possible trend could simply result from C0302 and
H1543 being in or near groups, and thus having a larger ``dark-matter
floor'' then the other three lenses. However, this would be the
opposite trend to e.g. PG1115+080 that shows a steeper than isothermal
mass profile (TK02) and is also close to a group (Kundi\v{c} et
al. 1997). A possible mechanism to explain steepening of the mass
density profile of the lens through interaction with a group/cluster
could be tidal stripping (e.g. Natarajan et al.\ 2002; TK03) of the
outer halo. Thus, if the lens was located at the center of the
group/cluster, where a significant amount of cluster dark matter is
present the observed profile could be flatter than isothermal because
of the ``dark-matter floor''. Viceversa, if the lens had been through
the cluster/group center, deep enough to experience tidal stripping,
but was observed on an outbound orbit far enough from the pericenter,
the ``dark-matter floor'' would no longer be relevant and a steeper
profile would be observed.

\begin{figure*}[t]
\begin{center}
\leavevmode \hbox{ \epsfxsize=0.95\hsize \epsffile{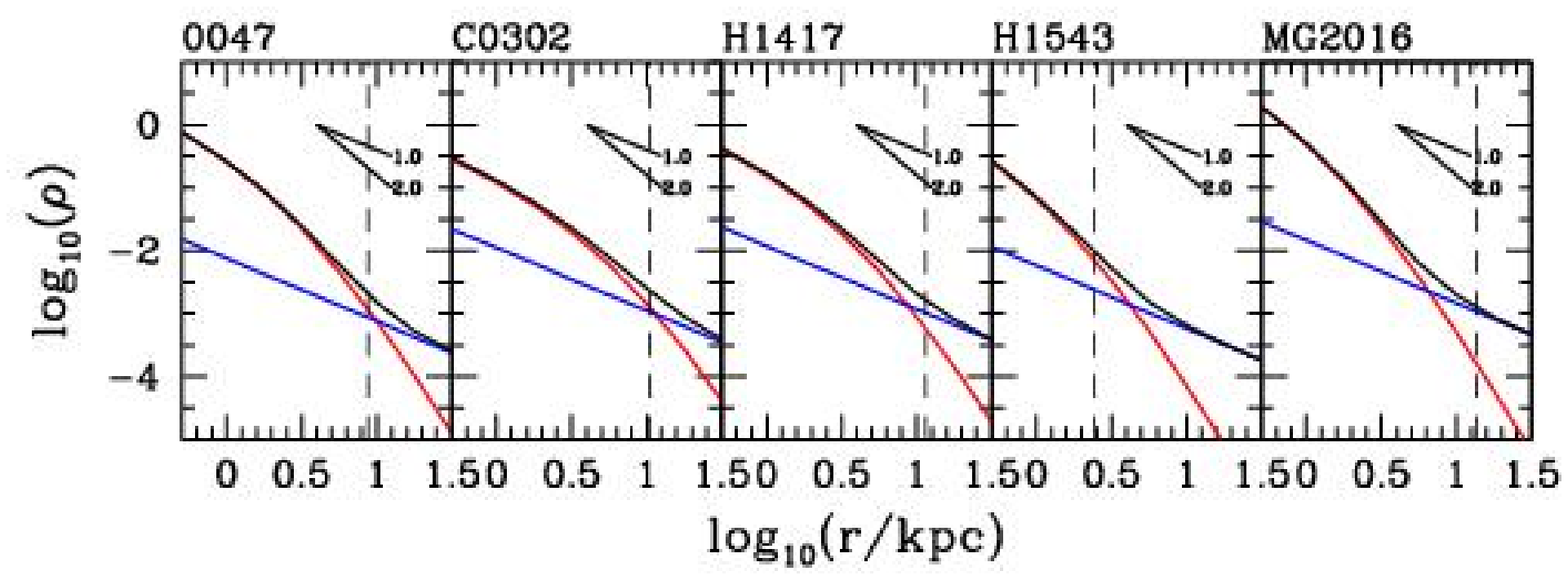}
} \figcaption{\label{fig:profiles} Density profiles of the five high-z
LSD lens systems, representing the most likely isotropic
($r_i=\infty$) model for a dark-matter halo (blue line) with
$\gamma=1$ ($\rho_{\rm DM}\propto r^{-\gamma}$) and a Hernquist
stellar mass density profile (red line). The total mass density
profile is plotted as a solid black line. The density is in units of
$10^{10}$~M$_\odot$\,kpc$^{-3}$ and the dashed line indicates the
Einstein radius. The two slopes for $\gamma'=1$ and 2 are indicated
for reference.}
\end{center}
\end{figure*}

Second, could the internal structure of luminous and dark matter in
these E/S0 galaxies alone explain a trend or large scatter? In the
local Universe, we know that `rotation' curves of early-type galaxies
show quite a variety of slopes within 1--2 effective radii
(e.g. Bertin et al. 1994; Gerhard et al. 2001; Romanowsky et
al. 2003), from rising ($\gamma'<2$) to declining ($\gamma'>2$). We
might expect that the average and scatter of $\gamma'$ would depend on
the ratio between the Einstein radius and the effective radius and the
fraction of dark matter $f_{\rm DM}$ contributing to the mass inside
the Einstein radius. The larger this ratio, the smaller the effect of
baryons, and the more $\gamma'$ will be a probe of the dark-matter
effective slope (possibly including some effects of nearby clusters or
groups). In this case $\gamma'$ will increase in three stages (see
Fig.~\ref{fig:profiles}): First, for $r<r_*$ (see Section~5) the slope is
typically dominated by luminous mass and will have $\gamma'<2$ for an
\dvp\ profile. Second, for $r>r_*$ a transition takes place where dark
matter becomes more prominent and the combination of luminous plus
dark matter add to an effective slope of $\gamma'\approx 2$. However,
in the case that the fraction of dark matter ($f_{\rm DM}$) inside
$R_{\rm Einst}$ is relatively small, one can expect that a very rapid
transition can occur to $\gamma'> 2$ (e.g. PG1115+080), if the luminous
component remains dominant at $r\ga r_*$ for $R_{\rm Einst}$ somewhat
larger than $r_*$. Clearly this is very sensitive to $f_{\rm DM}$ and
$R_{\rm Einst}/R_e$ and small variations in their value can induce
large fluctuations in $\gamma'$ for lenses with $R_{\rm Einst}/R_e \la
1$. For larger values of $R_{\rm Einst}/R_e$, the larger radial range
covered will result in less scatter for the same changes in $f_{\rm
DM}$. Third, around the break radius, a transition is expected from
the region where $\gamma'\approx 2$ to a dark-matter dominated regime
where $\gamma'\approx 3$ (see e.g.\ Seljak 2002 and Kneib et al.\ 2003
for a discussion of mass distribution at large radii from weak-lensing
studies).

It seems to us that most likely a combination of the effects discussed
above is required to interpret the observed trends and scatter in
$\gamma'$. Group and cluster halos exist and must necessarily play a
role (e.g.\ Lehar et al. 1997; Kundi\v{c} et al. 1997a\&b; Tonry 1998;
Tonry \& Kochanek 1999, 2000; Blandford, Kundi\v{c} \& Surpi 2001; Keeton,
Christlein \& Zabludoff 2000; Fassnacht \& Lubin 2002; Johnston et
al. 2003). However, a model where galaxies are simply isothermal and
appear more shallow if embedded in a group/cluster halo, is not
sufficient to explain the observations. Some degree of internal
scatter in the properties of the dark-matter halos of early-type
galaxies is required. This scatter could be the result of complex and
hierarchical formation history and baryonic cooling history, and/or
could be related to environmental effects (TK02b; Treu et al. 2003;
Natarajan, Kneib, \& Smail 2002).

Ultimately, whatever the underlying cause or interpretation, we can not
escape the conclusion that the inner total mass profile of E/S0
galaxies at $z\approx 0.5-1.0$ is {\sl on average} close to or
slightly more shallow than isothermal {\sl but also} that there is a
significant intrinsic r.m.s. scatter in $\gamma'$ of up to $\sim$0.3
(i.e.\ $\sim$15\% in density slope or $\sim$30\% in surface density slope).

\subsection{Lensing implications of inhomogeneity}

\label{ssec:imply}

In the previous subsection we concluded that E/S0 lens galaxies are
close to isothermal on average, but not quite, and that there is a
significant intrinsic scatter in the power-law slope of their total
inner mass profiles. Here, we will briefly discuss the consequences of
our finding on three important applications of gravitational lensing:
(i) The determination of the Hubble constant from gravitational time
delays; (ii) The determination of the star formation history of E/S0
galaxies from image-separation estimates of the FP; (iii) The
determination of the cosmological parameters from lens statistics.

Let us first consider the determination of the Hubble Constant from
gravitational time delays and consider a lens with unknown mass
profile, modeled as a singular isothermal ellipsoids. The observed rms
scatter in $\gamma'$ of $\sim$0.3 translates into a systematic
uncertainty on H$_0$ of $\sim$30\%, in addition to other
uncertainties. For example, an rms of $\sim$20\,\kmsmpc\ can be
expected if the true value is,say, H$_0$=65\,\kmsmpc. This range of
45--85 \kmsmpc\ covers the vast majority of determinations of H$_0$
that assume isothermal mass profiles (e.g.\ Schechter et al. 1997;
Impey et al. 1998; Biggs et al. 1999; Koopmans \& Fassnacht 1999;
Kochanek 2002a\&b; Koopmans et al. 2003; Wucknitz, Biggs \& Browne
2003) and could therefore in principle explain the mutual
inconsistency between the inferred values of H$_0$. Different samples
based on only a few lens systems could therefore lead to completely
different conclusions (e.g.\ Koopmans \& Fassnacht 1999; Kochanek
2002a\&b), if $\gamma'$ is not determined for each system directly.

In TK02b and K03, we applied the lensing and dynamics analysis
described above to two lens systems with measured time delays, finding
power-law slopes of $\gamma'=2.35\pm0.1\pm0.1$ for PG1115+080 and
$\gamma'=2.03\pm0.1\pm0.1$ for B1608+656, respectively, which lead to
values of H$_0$ of $59^{+12}_{-7}\pm3$~\kmsmpc\ and
$75^{+7}_{-6}\pm4$~\kmsmpc. In the case of PG1115+080, a $35\%$
increase was found for the value of H$_0$ from that expected from an
isothermal model with $\gamma'=2$ (i.e.\ H$_0$=44\,\kmsmpc; Impey et
al. 1998). A difference in slope of 0.3 between these systems is
fully consistent with the r.m.s. scatter in $\gamma'$ that we find in
our sample and thus `in hindsight' may not totally be unexpected. 

Indeed, in this paper we have presented three lens galaxies with
deviations of $\gamma'$ in the opposite direction. Those systems would
lead to severe overestimates of H$_0$ if they were assumed to be
isothermal. As we have stressed before, one can overcome these
problems by directly measuring the mass profile with a combination of
lensing and stellar dynamics, or other methods (e.g.\ Wucknitz
2003). The statistical approach (i.e.\ lens galaxies are ``on
average'' isothermal) is not (yet) satisfactory, since the average
value of $\gamma'$ and its scatter are poorly determined at present.
For example, if the current samples of about 4--5 lenses (e.g.\
Koopmans \& Fassnacht 1999; Kochanek 2002a\&b) were drawn from a
Gaussian distribution of slopes with $\langle \gamma' \rangle \equiv2$
and a 1--$\sigma$ width of 0.3, one would most likely find many
`outliers' (based on errors that do not incorporate the uncertainty in
the mass profile) with very low or high inferred values of H$_0$. The
distribution of $\gamma'$ for a sample of lenses might also depend on
unknown selection functions and it appears therefore preferable to
obtain a direct measurement of the mass slope ($\gamma'$) for lenses
with time-delays.

Let us now turn our attention to the star formation history of
early-type galaxies. Can we use multiple image separation to estimate
the central stellar velocity dispersion of E/S0 galaxies to construct
a Fundamental Plane of lens galaxies and measure the evolution of
their mass-to-light ratio (Kochanek et al. 2000; Rusin et al. 2003;
van de Ven et al. 2003)? What is the accuracy of this approximation?
Based on our sample, we find that $\langle f_{\rm SIE}
\rangle<1$. Hence, if we had used the isothermal model to determine
the central velocity dispersion of our lenses from image separation,
we would have overestimated their effective mass ($\sigma^2\,R_{\rm
e}$) and underestimated the evolution of $M_*/L_B$. If $\langle f_{\rm
SIE} \rangle$ was redshift independent, it would cancel out by fitting
simultaneously the local intercept (as done by R03), impacting the
measurement only through the increased uncertainty due to the intrisic
scatter in $\langle f_{\rm SIE} \rangle$.  However, this could be a
redshift dependent effect, since, for example, the ratio of the
Einstein radius to the effective radius could depend on the redshift
of the lens, and therefore it could mimic evolutionary trends. Hence,
if the results from our sample of five E/S0 lens galaxies hold for the
larger sample of lens system, it could explain why most direct
measurements (e.g.\ Treu et al. 2002; Gebhardt et al. 2003; van der
Wel et al. 2004; this paper) indicate a slightly faster evolution of
the Fundamental Plane of E/S0 galaxies with redshift than those based
on lens-estimates of $\sigma$ (e.g.\ Rusin et al. 2003; van de Ven et
al.\ 2003). If this difference -- at the moment only marginally
significant -- was confirmed by more precise measurements, it would be
interesting to reverse the argument. If the lens-based estimate of the
evolution of the FP is slower than the direct measurement, then the
power-law slope of lens galaxies is on average flatter than isothermal
and $\langle f_{\rm SIE}\rangle <1$. The difference in the
evolutionary rates would provide another measurement of $\langle
f_{\rm SIE}\rangle$ and more importantly on $\langle \gamma' \rangle$.

Finally, we shortly discuss the effects on lensing statistics. If
galaxies have $\langle f_{\rm SIE}\rangle < 1$ and $\gamma'=2$, then
statistical models that use isothermal lens mass models with
$\sigma_{\rm SIE}=\sigma$ will tend to underestimate the lensing
cross-section of a population of lenses, since the correct
$\sigma_{\rm SIE}=\sigma/f_{\rm SIE}>\sigma$.  However, since we have
found that $f_{\rm SIE}< 1$ for many lenses might actually be
associated with $\gamma' < 2$, the effect is not so clear, since
galaxies with a more shallow mass profiles have lower lens
cross-sections for a fixed enclosed mass. On the other hand, they are
more massive when normalized to the same $\sigma$ (i.e.\ a shallower
profile lowers $\sigma$ for a fixed mass, hence to increase it to the
same value as for a steeper profile, its mass needs to be
increased). We are generally in the latter situation, since
statistical models are typically normalized to an observed
distribution function of $\sigma$. It becomes even more complex,
however, since all these effects need to be integrated over a
distribution function of $\gamma'$ and effects of the magnification
bias as function of mass profile need to be accounted for.  The full
treatment of this problem goes beyond the scope of this paper and is
left for future research.

To summarize, the main conclusion of this section is that a simple
isothermal model is probably not appropriate for precision
measurements, especially when only small samples are available, as it
is typically the case. The observed scatter does not seem surprising,
given, for example, the wide range of velocity dispersion profiles
observed for early-type galaxies, from declining to increasing with
radius (see e.g., Gerhard et al.\ 2001; Kelson et al.\
2002). Unfortunately the samples are still too small and the selection
functions too poorly characterized to find out what is the {\it
distribution} of total mass density profiles and whether the {\it
distribution} for lens samples differs from the one from
morphologically or X-ray selected samples. As far as the lenses are
concerned, at this stage the average slope appears to be slightly
shallower than isothermal, in marginal contrast with other
determinations, using different methods and samples (e.g. Rusin et
al. 2003). Although the difference is relevant for precision
measurements (e.g. of H$_0$) based on statisical arguments, we believe
that what is remarkable is the relative agreement on the peak of the
distribution being around $\gamma'=2$, given the small size of the
samples and the unknown sample selection biases. A larger number of
lenses with precisely measured mass density profiles is needed to make
further progress on these issues.

\section{Summary}

\label{sec:sum}

We have presented new spectroscopic measurements for three
gravitational lens systems C0302 ($z=0.938$), H1543 ($z=0.497$), and
H1417 ($z=0.81$) as part of the Lenses Structure \& Dynamics (LSD)
Survey. Long integrations with ESI at the Keck--II Telescope have
yielded extended stellar velocity dispersion profiles of all three
lens E/S0 galaxies, extended approximately to the effective radius. A
blue spectrum taken with LRIS--B has revealed the redshift of the
lensed arc in H1543 ($z=2.092$). Together with two previously
published systems, MG2016 ($z=1.004$; TK02; see also Koopmans et al.\
2002) and 0047 ($z=0.485$; KT03), this paper presents the analysis of
the current high-redshift sample ($z\approx0.5-1.0$), consisting of
five pressure-supported E/S0 galaxies.

The spectroscopic data have been combined with surface photometry from
Hubble Space Telescope (HST) archival images to study the evolution of
the stellar populations via the evolution of the intercept of the
Fundamental Plane (FP). For the sample of five LSD lenses we find
$d \log (M/L_{B}) / d z = -0.72\pm0.10$, i.e.\
1.80$\pm$0.25 magnitudes of dimming between $z=1$ and $z=0$. In a pure
luminosity evolution scenario, this measurement can be interpreted as
the results of a relatively young luminosity-weighted age of the
stellar populations. A scenario were most of the stars were formed at
high redshift ($>2$) while a small fraction of stars ($\sim10$\%) is
formed in secondary bursts between $z=1$ and $z=0$, provides a simple
explanation for this result, as well as several independent pieces of
evidence (evolution of the Luminosity Function of E/S0 galaxies;
spectroscopic evidence of recent minor episodes of star formation in
distant E/S0; properties of local E/S0; see the discussion in
Section~\ref{sec:FP})

New gravitational lens models of the three systems C0302, H1417 and
H1543 have been presented. H1417 has the classical `quad' morphology
and can be successfully modeled with a Singular Isothermal Ellipsoid
(SIE) mass distribution, with an external shear and a gradient in the
local convergence that roughly aligns with the galaxy major axis and
which we interpret as an internal asymmetry in the galaxy. In
contrast, the lens systems C0302 and H1543 are characterized by a
source lensed into extended arc-like features. These systems are
modeled with an algorithm that allows for a non-parametric image
reconstruction (see Appendix), incorporating some of the techniques by
Wallington et al. (1996) and Warren \& Dye (2003). Both lenses are
successfully modeled with a SIE mass model with external shear. In the
case of H1543, a nearby ($4\farcs7$) massive galaxy at the same
redshift as the main lens is included in the lens model as a Singular
Isothermal Sphere (SIS). The Einstein radii ($R_{\rm Einst}$), SIE
velocity dispersions ($\sigma_{\rm SIE}$) and enclosed masses (M$_{\rm
Einst}$) of the three lenses are used to perform a joint lensing and
dynamical analysis, with the following results.

\begin{enumerate}

\item Constant mass-to-light ratio models (i.e.\ mass follows light)
are rejected at better than 99\% CL for all five E/S0 lens galaxies. A
dark-matter halo with a mass density profile flatter than the luminous
component is needed in all cases. The fraction of dark matter inside
the Einstein Radius ($f_{\rm DM}$) is 37--72\% (isotropic stellar
orbits) and 15--65\% inside the effective radius.

\item Modeling the total mass density profile of the lenses as a
single power law density distribution $\rho_{\rm tot}\propto
r^{-\gamma'}$, the effective slope $\gamma'$ is found to be on average
somewhat smaller than isothermal, i.e. $\langle \gamma'\rangle =1.75$
with and rms scatter of 0.20 (for isotropic velocity ellipsoid;
$\langle\gamma'\rangle=1.57$ with rms 0.35 for radial anisotropy) for
our sample of five lenses. If we include the two others systems that
have a similar analysis, these values increase by $\sim$0.15 and the
rms increases to $\sim$0.30. Consistent with these findings, the ratio
$f_{\rm SIE}=\sigma/\sigma_{\rm SIE}$ between central velocity
dispersion and velocity dispersion of the Singular Isothermal
Ellipsoid (SIE) mass model that best fits the lensing constraints is
$\langle f_{\rm SIE}\rangle =0.87$ with an rms scatter of 0.08.

\item The average mass-to-light ratio of the luminous component
$\langle M_*/L_B \rangle =(2.3\pm0.6)\,h_{65}$\,\mlu\ is smaller than
the average value for early-type galaxies in the local Universe
$(7.3\pm2.1)\,h_{65}$\,\mlu, consistent with passive evolution of a
relatively old stellar population. The stellar mass to light ratio
obtained from the joint lensing and dynamics analysis is found to
evolve as $\langle d \log (M/L_{\rm B}) /dz\rangle =-0.75\pm0.17$ in
excellent agreement with the independent measurement obtained via the
Fundamental Plane.

\item The most precise constraints to date are obtained on the inner
slope of the dark-matter halo $\gamma$. We find the following 68\%
confidence limits: $\gamma<0.58$ (anisotropic velocity ellipsoid with
$r_i=R_e$) and $0.93<\gamma<1.48$ or $\gamma=1.3^{+0.2}_{-0.4}$ (68\%
CL) for $r_i=\infty$. Thus, our data are consistent with CDM numerical
simulations (with $\gamma$=1--1.5) only if the velocity ellipsoid is
not significantly radially anisotropic and baryonic collapse, during
galaxy formation, did not significantly steepen the mass density
profile as would be expected in simple adiabatic contraction scenarios
(Blumenthal et al.\ 1986; Mo, Mao \& White\ 1998; Keeton 2001;
Kochanek\ 2002).

\end{enumerate}

\section{Conclusions}

\label{sec:conc}

In conclusion, the following picture seems to be emerging from the
analysis of the complete high-redshift LSD sample. High redshift
early-type galaxies are approximately isothermal ellipsoids, but not
exactly.  Our current sample seems to indicate that on average the
effective slope of the mass density profile inside the Einstein radius
might be slightly smaller than 2, i.e. total mass density profile
flatter than isothermal, albeit $\gamma'=2$ is generally within the
range of the distribution (e.g.\ the first two objects we analyzed
were almost exactly isothermal, TK02, KT03). A possible cause for
departure from homogeneity could be the environment of early-type
galaxies.  Contributions from a relatively flat group or cluster
dark-matter halo could introduce a ``floor'' of mass at the position
of the lens causing the total mass density profile to appear
effectively flatter. Independent external evidence (such as the
presence of nearby galaxies at the same redshift) indicates that
possibly all three lenses for which we found $\gamma'<2$ might be
members of a group, and therefore this mechanism would appear to be a
likely explanation, at least for this sample. However, this mechanisms
does not appear to be sufficient to account for all the observed
scatter, because most of the scatter is observed for $R_{\rm
Einst}/R_{\rm e}<1$, where the galaxy mass distribution should
dominate over a group halo, and because there are cases where
$\gamma'>2$ is observed (such as PG1115+080 and B1608+656; the former
also has a nearby compact group). We therefore conclude that the
scatter in the total mass density profile is associated with intrinsic
scatter in the ratio of dark matter to luminous matter in the inner
regions of high redshift early-type galaxies, similar to what is
observed in the local Universe (Bertin \& Stiavelli 1993; Bertin et
al.\ 1994; Gerhard et al. 2001).

Regardless of the physical interpretation, the observed scatter in
the effective slope (i.e. 10--15\% in density slope) still implies a
remarkable degree of structural homogeneity between early-type
galaxies from a galaxy formation point of view. Whatever the details
of the mass assembly and star formation history, E/S0 galaxies end up
being close to isothermal, for example expected in (incomplete)
violent relaxation scenarios (Lynden-Bell 1967; Shu 1978; van Albada
1982; Bertin \& Stiavelli 1993; Hjorth \& Madsen 1991; 1995). From an
evolutionary point of view, the intrinsic scatter in the mass density
profiles of high redshift E/S0 galaxies does not seem to be much
different from that of local E/S0 galaxies, providing no evidence for
much structural evolution within the last 4--8 Gyrs.

Although the homogeneity is remarkable from a galaxy-formation point
of view, the observed scatter in $\gamma'$ is large enough that the
isothermal approximation might not be good for some applications,
particularly when they depend critically on the mass slope. Meaningful
examples include the determination of $H_0$ from gravitational
time-delays, where a scatter of $\sim$0.3 in $\gamma'$ translates into
a scatter of $\sim$30\% in $H_0$ for a given time delay (Saha 2000;
Wucknitz 2002; Kochanek 2002; TK02b). Thus, it appears necessary to
use external information -- such as internal kinematics -- to pinpoint
$\gamma'$ and $H_0$ to a level of accuracy (10-15\%) competitive with
other methods (Koopmans et al.\ 2003; see also Wucknitz et al.\
2003). A precision measurement based on statistical assumptions of
$\gamma'$ will probably have to wait until the distribution of
$\gamma'$ is observationally well characterized, and a large enough
sample of lenses with time delays is available to minimize variance.

The other main result of this paper is the decomposition of the total
mass distribution into a luminous component and a dark-matter halo.
High-redshift early-type galaxies are inconsistent with constant
mass-to-light ratio mass models. The mass-to-light ratio has to
increase significantly with radius, consistent with the presence of
dark-matter halos flatter than the luminous component (dark matter
makes up a substantial fraction of the mass of the lenses inside the
Einstein Radius, of order 40--70\%).

The mass-to-light ratio of the luminous component is smaller than in
the local Universe, consistent with passive evolution of a relatively
old stellar population. The agreement between the evolution of the
stellar mass to light ratio measured directly and that determined from
the evolution of the intercept of the Fundamental Plane is another
argument against strong structural and dynamical evolution. If
early-type galaxies had changed their mass distribution significantly
between $z\sim1$ and today, they would have to be doing so while
preserving the mapping between velocity dispersion, radius and stellar
mass. A simpler -- and to our eyes preferable -- explanation would be
one involving no or little dynamical evolution.

The precise mass density profile of the dark-matter halos is harder to
constrain, since most of the mass at small radii is
luminous. Nevertheless, our sample of five lenses allows us for the
first time to set limits on the inner dark-matter density slope
$\gamma$ that are interesting to compare with cosmological
simulations. Our measurement can be reconciled with numerical
cosmological only if (i) the velocity ellipsoid is not significantly
radially anisotropic; and (ii) the collapse of baryons to form the
galaxy did not steepen the dark-matter halo more than a few tenths in
$\gamma$. The latter constraint is clearly inconsistent with simple
adiabatic collapse models (Blumenthal et al.\ 1986) and suggests that
different mechanisms are involved in the accretion of stars of the
centers of the halos of early-type galaxies. This is further supported
by lensing statistics results, which imply a low value of $f_{\rm
DM}(<R_e)<0.33$ (95\% CL; Keeton 2001). Assuming adiabatic contraction
implies that the initial slope ($\gamma_i$) of the dark-matter halo
(i.e.\ before contraction) was shallower than predicted from
$\Lambda$CDM models (i.e. $\gamma_i<1$) and introduces an additional
inconsistency. Mechanisms such as those proposed for clusters (El-Zant
et al.\ 2003; Nipoti et al.\ 2003), where stars form first in
satellites, which are then accreted as dissipationless particles could
provide the desired effect (see also Loeb \& Peebles 2003), although a
detailed comparison of data with theory will have to wait until
cosmological simulation including a realistic treatment of star
formation become available.

\acknowledgments We thank Eric Agol, Andrew Benson, Giuseppe Bertin,
Roger Blandford, Stefano Casertano, Luca Ciotti, Richard Ellis, Chris
Fassnacht, Jean-Paul Kneib, Chris Kochanek, David Rusin, and Massimo
Stiavelli for useful comments on the LSD project and stimulating
conversations. The anonymous referee's prompt and insightful comments
improved this paper. We are grateful to Myungshim Im for personally
suggesting to target H1543. We thank the Caltech TAC for the generous
allocations of Keck observing time that made this program
feasible. The use of the Gauss-Hermite Pixel Fitting Software and
Gauss-Hermite Fourier Fitting Software developed by R.~P.~van der
Marel and M.~Franx is gratefully acknowledged. The ESI data were
reduced using software developed in collaboration with D.J.~Sand.  We
acknowledge the use of the HST data collected by the CASTLES
collaboration. TT acknowledges support from NASA through Hubble
Fellowship grant HF-01167.01.  LVEK acknowledges support from Space
Telescope Science Institute through a Institute Fellowship. LVEK and
TT acknowledge support by an archival research grant from NASA
(STScI-AR-09960). We thank the people that developed ESI and LRIS, for
building such wonderful instruments. We are very grateful to the staff
of the Keck Observatory for their invaluable help in making the most
out of our observing runs. Finally, the authors wish to recognize and
acknowledge the very significant cultural role and reverence that the
summit of Mauna Kea has always had within the indigenous Hawaiian
community.  We are most fortunate to have the opportunity to conduct
observations from this mountain.

\appendix

\section{Delensing of images on a grid}

Suppose we have a lensed image, $\dv$, on a grid (e.g.\ a CCD
image)\footnote{All two-dimensional grids are represented as vectors
in which consequative rows are placed behind each other.} in which all
emission not associated with the lensed source has been masked and/or
subtracted. Hence the grid should be a noisy, blurred and lensed
representation of the true underlying source brightness distribution,
which it is our aim to reconstruct. Suppose also that we can construct
a lens-operator $\lo$ (depending on the parameters of our lens model)
which acting on a source grid, $\sv$, produces a lensed image of the
source. Suppose further that a blurring operator, $\bo$, exists which
acting on $\lo$\,$\sv$ produces a blurred lensed image. Putting this
together, we have
\begin{equation}
	{\bf B\,L\,\hat{s} = \hat{d_t} + \hat{n}},
\end{equation}
where $\nv$ represents the noise in the observed image and ${\bf
\hat{d_t}}=\dv-\nv$ is the noise-free lensed image.  Note that the
size or shape of the image and source grids and their pixel sizes are
irrelevant to the this problem.  Furthermore, neither the grid nor the
pixels have have to be rectangular and connected (e.g. the image grid
could have a gap). As long as the value of each observed pixel can be
written as a linear combination of source pixel values, the above
equation is applicable. For simplicity, however, we will assume
rectangular grids and pixels for the remainder of this paper.
Naively, one would think that the solution to this problem, i.e.\ the
source brightness distribution, can easily be found by inverting the
above equation through $\sv=({\bf B\,L})^{-1}\,\dv$. However, this is
a notoriously ill-posed problem and noise in the observed image will
typically lead to unacceptably poor reconstructions. We will not
further discuss this solution.
The problem can be regularized, however, suppressing the effects of
$\nv$ on the final source reconstruction (Thikonov 1965).
Mathematically, we would like to find the source grid $\sv$ and the
parameters of the lens model (i.e. \lo) that minimize the following
equation
\begin{equation}
     C(\lambda)=||{\bf B\,L\,\hat{s}-\hat{d}}||^2_2 + \lambda ||{\bf
     H\,\hat{s}}||_2^2,
\end{equation} 
where $\ro$ is a regularization operator and $\lambda$ determines the
weight given to the regularization term. For simplicity we have also
written $(\bo\,\lo)/\nv \rightarrow (\bo\,\lo)$ and ${\bf \hat{d}}/\nv
\rightarrow {\bf \hat{d}}$ (see Press et al.\ 1992). The first term in
Eq.(2) is simply the $\chi^2$ term, whereas the second term regulates
the ``smoothness'' of the final solution. In particular, if $\ro={\bf
I}$, then the regularization term is simply the sum over the squared
pixel values in the source grid. The latter term, however, can also
regulate the smoothness of the derivatives in the source grid, or be
replaced by a maximum likelihood or maximum entropy term (e.g. Whyth
et al. 2003).
The solution to Eq.(2) is the solution to the set of linear equations
\begin{equation}
       \left[(\bo\,\lo)^{\bf T} (\bo\,\lo) + \lambda\,\ro\right]\,\sv
	= (\bo\,\lo)^{\bf T} \, \dv.
\end{equation}
This equation has a unique solution, thanks to the regularization
term, and can be solved through standard techniques and using freely
available linear algebra packages for large sparse matrices.

\section{The Lensing \& Blurring Operators}

The main problem that is faced in constructing a lens operator is the
fact that an image pixel, when projected on the source grid, in
general will not exactly coincide with a source pixel. One way to
solve this (Warren \& Dye 2003) is to determine the lensed image for a
given source pixel of unit flux, blurred by the PSF and regridded to
the image grid. This is repeated for all source pixels. The source is
then reconstructed by finding the set of source-pixel weights (i.e.\
fluxes for each source pixel) and the corresponding lensed images, for
which their linear combination best reconstructs the observed
image. This requires inverting the lens equation and is numerically
expensive for complex models. It also requires finding all lensed
images for each pixel. There are no simple algorithms that guarantee
this for complex mass models. 
However, one can also construct a lens operator that does not require
the lens inversion. Before describing this, we first introduce some
definitions. We assume that the source grid has $K\times L$ pixels ($K$
columns and $L$ rows) and similarly, the (as of yet unblurred) image
grid, $\dv'$, has $M\times N$ pixels. The values of each pixel are
$s_{k,l}$ and $d'_{m,n}$, respectively. Hence $\sv=\{s_{i=k+(l-1)K}\}$
with $i=1...KL$ with $k=1...K$ and $l=1...L$. Similarly,
$\dv'=\lo\sv=\{d'_{j=m+(n-1)M}\}$ with $j=1...MN$ with $m=1...M$ and
$n=1...N$.
It is now very easy to contruct a particular implementation of the
lens operator, $\lo$, which is a matrix of size $KL \times MN$ and
entries $l_{i,j}$. We emphasize that our choice is not unique, but it
is simple and fast.
First, each pixel $j=m+(n-1)M$ is cast back on the source plane to a
position $\vec{y}_j$, using the lens equation. If the position is
outside the (pre-defined) grid, one simply continues to the next
pixel. In general, however, the lensed image grid and source grid will
be defined to overlap as much as possible to reduce redundancy,
although this is not a requirement.  Second, one determines the four
pixels in the source grid that enclose $\vec{y}_j$, say
$(r+\mu,s+\nu)$ for $\mu=0,1$ and $\nu=0,1$.  Suppose that the
position of $(r,s)$ is $\vec{y}_{r,s}$ and $(t,u)=\vec{y}_j -
\vec{y}_{r,s}$. The flux of pixel $j$ can then be written as a linear
combination of the four source pixel fluxes
\begin{equation}
	d'_j = \sum_{\mu=0}^{1} \sum_{\nu=0}^{1} w_{\mu,\nu} \,
	        s_{i=(r+\mu)+(s+\nu-1)K}
\end{equation}
where
\begin{eqnarray}
	w_{0,0}&=&(1-t)(1-u) \nonumber\\
	w_{1,0}&=&t(1-u)     \nonumber\\
	w_{0,1}&=&tu         \nonumber\\
	w_{1,1}&=&(1-t)u. 
\end{eqnarray}
This is simply a bilinear interpolation of the four source pixel
fluxes, but more complicated linear schemes can be constructed,
although they will generate more entries in the lens operator (except
if ones chooses to use an enclosing triangle).
The four entries in the lens operator, at $l_{i,j}$, for image pixel
$j$ are then the values of $w_{\mu,\nu}$ at $i=(r+\mu)+(s+\nu-1)K$ for
$\mu=0,1$ and $\nu=0,1$. Finally, $\lo$ contains at most $4MN$
entries.\footnote{Note that $\lo$ contains a fraction of $\le4/(KL)$
non-zero elements. For a 100$\times$100 image grid, this is
$\le$0.04\%. It is obvious that sparse-matrix packages are required,
or otherwise $\lo$ alone would require $\sim$1\,GB of computer memory to
store.} 

\medskip

The next step is to construct the blurring operator, $\bo$, that acts
on the lensed image $\dv'$. As illustration, suppose that the PSF is a
square grid of size $(2H+1)\times(2H+1)$ pixels with values
$p_{\mu,\nu}$ with $\mu=-H...H$ and $\nu=-H...H$ and peaks at
$p_{0,0}$. The sum of the PSF pixels adds to unity.
The entries in the blurring operator, $b_{g,h}$ are then simply the
values of $p_{\mu,\nu}$ at $g=(h+\mu)+(h+\nu-1)M$, if and only if
$1\le (h+\mu)\le M$ and $1 \le (h+\nu)\le N$ for each $h=1...MN$. 
Notice here that this method allows one to define a color-dependent
blurring operator. In that case, for each $h$ (i.e.\ pixel in the
image plane), one can use a PSF with values $p_{\mu,\nu, h}$ that
depend on the local color of the pixel. 
We also note that an
extinction correction can be done by setting the integral over the
PSF, $\Sigma_\mu \Sigma_\nu p_{\mu,\nu} = e^{-\tau} <1$, where $\tau$
is the optical depth due to extinction. (Note that this requires a
color and extinction model, since both effects occur {\sl before}
blurring.)

\section{Practicalities in the Optimization}

The pixel size in the source plane is set roughly by the largest pixel
magnification in the image plane. If the source-plane pixels are
chosen too large, the resulting image shows the effects of the mapping
of individual source-plane pixels. If they are chosen too small,
however, the source breaks up in ``strings'' of pixels that map
closely to the image-plane pixels, but no flux in between (e.g.\ for
$\bf H$=$\bf I$) Mathematically, these solutions are equivalent, but
physically clearly not. We therefore set the pixel size such that the
source does not tend to break up and adjacent image-plane pixels
roughly map onto adjacent source-plane pixels. We then set $\lambda$
to an initially large value of typically~$\sim$0.1 and minimize
$C(\lambda)$ by varying the lens-model parameters (see WD03 for
details). We then lower the value of $\lambda$ slowly, continuing to
optimize the model parameters, until a reduced $\chi^2$ of unity is
reached. In general we find that the resulting mass-model parameters
are relatively robust against changes in either the source-plane pixel
size and the value of $\lambda$, as long as the source remains
compact, and we refer to WD03 for a more thorough discussion of
different choices.

\end{document}